\documentclass[oldversion]{aa}
\usepackage[pdftex]{graphicx}
\usepackage{epsfig}
\usepackage{amsmath}
\usepackage{wasysym}
\usepackage{marvosym}
\usepackage{txfonts}
\pdfoutput=1
\usepackage[pdftex]{color,xcolor}
\usepackage[pdftex]{graphicx}
\usepackage[backref]{hyperref} 
\usepackage{hyperref}
\hypersetup{pdfauthor=Paul Molliere}
\hypersetup{backref=true, pagebackref=true, hyperindex=true, breaklinks=true,colorlinks=true,urlcolor=blue, linkcolor=blue,citecolor=blue,pagecolor=red, bookmarks=true, bookmarksopen=true}
\usepackage{epstopdf}
\hypersetup{backref=true}
\usepackage{lipsum}
\usepackage{natbib}
\usepackage{mathtools}

\bibpunct{(}{)}{;}{a}{}{,} 
\usepackage{mhchem}

\def\ptrad{\emph{{petitRADTRANS}}}
\def\ptdoc{\url{https://petitradtrans.readthedocs.io}}

\def\rsun{{\rm R}_\odot}

\def\f1{f_{\rm I}}
\def\mj{{\rm M}_{\textrm{\tiny \jupiter }}}

\newcommand{\rj}{{\rm R}_{\textrm{\tiny \jupiter}}}

\def\beq{\begin{equation}}
\def\eeq{\end{equation}}

\def\t2{\tau_{\rm II}}

\def\sigmas0{\Sigma_{\rm s,0}}

\def\petit{\emph{petitCODE}\ }

\def\({\left(}
\def\){\right)}
\def\<{\left<}
\def\>{\right>}

\def\tr{\mathcal{T}}

\newcommand{\rch}[1]{{{#1}}}
\newcommand{\rcht}[1]{{{#1}}}

\begin{document}

\title{\emph{petitRADTRANS}}
\subtitle{a Python radiative transfer package for exoplanet characterization and retrieval}

\author{P. Molli\`{e}re\inst{1,2}  \and J.P. Wardenier\inst{1}  \and R. van Boekel\inst{2} \and Th. Henning\inst{2} \and K. Molaverdikhani\inst{2} \and I. A. G. Snellen\inst{1}}

\institute{Leiden Observatory, Leiden University, Postbus 9513, 2300 RA Leiden, The Netherlands \and Max-Planck-Institut f\"ur Astronomie, K\"onigstuhl 17, 69117 Heidelberg, Germany}

\offprints{Paul MOLLIERE, \email{molliere@strw.leidenuniv.nl}}

\date{Received -- / Accepted --}

\abstract{We present the easy-to-use, publicly available, Python package \ptrad, built for the spectral characterization of exoplanet atmospheres. The code is fast, accurate, and versatile; it can calculate both transmission and emission spectra within a few seconds at low resolution ($\lambda/\Delta\lambda$ = 1000; correlated-k method) and high resolution ($\lambda/\Delta\lambda = 10^6$; line-by-line method), using only a few lines of input instruction.  The somewhat slower, correlated-k method is used at low resolution because it is more accurate than methods such as opacity sampling. Clouds can be included and treated using wavelength-dependent power law opacities, or by using optical constants of real condensates, specifying either the cloud particle size, or the atmospheric mixing and particle settling strength. Opacities of amorphous or crystalline, spherical or irregularly-shaped cloud particles are available. \rch{The line opacity database spans temperatures between 80 and 3000~K, allowing to model fluxes of objects such as terrestrial planets, super-Earths, Neptunes, or hot Jupiters}\rcht{, if their atmospheres are hydrogen-dominated}\rch{. Higher temperature points and species will be added in the future, allowing to also model the class of ultra hot-Jupiters, with} \rcht{equilibrium temperatures} \rch{$T_{\rm eq} \gtrsim 2000$~K.} Radiative transfer results were tested by cross-verifying the low- and high-resolution implementation of \ptrad, and benchmarked with the \emph{petitCODE}, which itself is also benchmarked to the \emph{ATMO} and \emph{Exo-REM} codes. We successfully carried out test retrievals of synthetic JWST emission and transmission spectra \rch{(for the hot Jupiter TrES-4b, which has a $T_{\rm eq}$ of  $\sim$ 1800~K)}. The code is publicly available at \url{http://gitlab.com/mauricemolli/petitRADTRANS}, and its documentation can be found at \ptdoc.}

\titlerunning{The petitRADTRANS package for exoplanet retrieval}
\authorrunning{P. Molli\`ere et al.}

\maketitle

\section{Introduction}
The characterization of exoplanets via the interpretation of their spectra is an important and fast-developing branch of exoplanet science. Especially inferring the actual distribution of the parameters describing the atmospheres, so-called retrieval, is now a commonly carried out and described analysis. Parameters that are inferred are the atmospheric vertical (and sometimes horizontal) temperature and abundance structure, parameters defining the exoplanetary clouds, cloud coverage, or even the stellar parameters, such as spot or facula coverage \citep[see, e.g.,][for a non-exhaustive list]{madhusudhanseager2009,madhusudhan2011,leefletcher2012,bennekeseager2012,linezhang2012,linewolf2013,lineyung2013,lineknutson2014,leeirwin2014,benneke2015,lineteske2015,waldmannrocchetto2015,waldmanntinetti2015,greeneline2016,linemarley2016,lineparmentier2016,fengline2016,rocchettowaldmann2016,samlandmolliere2017,macdonaldmadhusudhan2017,laviemendoca2017,pinhasrackham2018,fisherheng2018,pinhasmadhusudhan2019}. This analysis method takes root in the characterization efforts of the solar system planets \citep[see, e.g.,][and the references therein]{irwinteanby2008}.

Retrievals usually consider observations at low resolution, with the most reliable data coming from space-based telescopes such as \emph{Hubble} and \emph{Spitzer}. The work by \citet{brogiline2016,brogiline2018} has recently advertized and demonstrated the possibility of carrying out retrieval analyses using ground-based, high-resolution spectra. This is particularly exciting because high-resolution data allows to infer additional planetary properties, such as wind speeds \citep[e.g.,][]{snellendekok2010,flowersbrogi2018}, spin rates \citep[e.g.,][]{snellenbrandl2014,schwarzginki2016,bryanbenneke2018}, or even atmospheric cloud maps \citep[][]{crossfieldbiller2014}. Moreover, detecting and measuring isotopologue abundance ratios in exoplanets may allow to probe planet formation and atmospheric evolution processes \citep{mollieresnellen2018}.

\begin{table*}[t!]

\centering
{ \footnotesize
\begin{tabular}{l|lll}
\hline \hline
Property &  {\bf petitRADTRANS} (this paper) & \petit \citep{mollierevanboekel2015,mollierevanboekel2016} \\ \hline
Temperature & Parametrized, e.g., \citet{guillot2010} & Radiative-convective equilibrium \\
Abundances & Parametrized, e.g., vertically constant & Chemical equilibrium, from elemental abundances \\
Scattering & Included for transmission spectra only & Transmission \& emission,  also during structure iteration \\
Cloud options & Power-law \& condensation clouds & Condensation clouds \\
Cloud particle size & $f_{\rm sed}$ and $K_{\rm zz}$, or parametrized & $f_{\rm sed}$ and $K_{\rm zz}$, or  parametrized\\
Particle size distribution & log-normal, width variable & log-normal, width variable \\
Cloud abundance & Parametrized & \citet{ackermanmarley2001} or \citet{mollierevanboekel2016} \\
Wavelength spacing $\lambda/\Delta\lambda$ & 1000 (correlated-k), 10$^6$ (line-by-line) & 10, 50, 1000 (correlated-k)\\
Validity transmission spectra$^{\rm (a)}$ & Clear \& cloudy cases & Clear \& cloudy cases \\
Validity emission spectra$^{\rm (a)}$ & Clear, from NIR wavelengths on ($\lambda >$ optical) & Clear \& cloudy cases \\
\hline
\end{tabular}
}
\caption{Summary and comparison of the general properties of \ptrad, the code presented in this paper, and \emph{petitCODE}, to which \ptrad \ spectral calculations are compared in Section \ref{sect:comp_low_low_res}. The $f_{\rm sed}$ and $K_{zz}$ parameters are the mass-averaged ratio of the cloud particle settling speed and the mixing velocity, and the atmospheric eddy diffusion coefficient, respectively, as described in \citet{ackermanmarley2001}. (a): \ptrad \ does not treat scattering for emission spectra, so emission spectra of \ptrad \ where cloud scattering is expected to play an important role should be used with caution.}
\label{tab:code_comp}

\end{table*}

Consequently, the application cases for retrieval and, more general, atmospheric characterization studies are manifold. The need for easy-to-use tools for exoplanet spectral synthesis that allow the user, observers and theorists alike, to build an intuition of the physical processes that shape exoplanet spectra, and to set up retrievals, is therefore evident. Ideally, such tools should be as versatile as possible, allow us to calculate spectra that are clear or cloudy, at high or low resolution, and calculate the planet's transmission or emission spectrum. Combining all of these capabilities in a single tool would be ideal, such that the user does not have to go on the hunt for a different code every time her or his model requirements change. Moreover, such a code should calculate spectra in a reasonable amount of time, to allow playing with the results interactively, and for carrying out retrievals. At the same time, the accuracy of the results should be reliable, and not sacrificed for computational speed. \emph{petitRADTRANS} is our attempt to fulfill the criteria stated above.
It is a code for spectral synthesis, to be used in exoplanet retrievals. This means that temperature and abundance profiles are required as free parameters, rather than being solved for in radiative-convective and chemical equilibrium. Consequently, temperature profiles, abundance and cloud parameters are the retrievable quantities. \ptrad \ is thus different from \emph{petitCODE} \citep{mollierevanboekel2015,mollierevanboekel2016}, which is our model for solving for atmospheric structures and spectra self-consistently. Also see Table \ref{tab:code_comp} for the differences in modeling philosophy between the two codes, the spectra of which will be compared in Section \ref{sect:comp_low_low_res}.


Before summarizing the capabilities of \ptrad, and the corresponding structure of this paper, we give a short overview of the tools and codes already publicly available: \emph{Tau-REx} \citep{waldmannrocchetto2015,waldmanntinetti2015} is a Python radiative transfer plus retrieval code which, similar to \ptrad, allows to calculate emission and transmission spectra, at low or high resolution, \rch{including clouds}. The code and some documentation are available on github\footnote{\url{https://github.com/ucl-exoplanets/TauREx_public}}.
The \emph{BART} code for exoplanet emission or transmission spectral retrieval is available from github\footnote{\url{https://github.com/exosports/BART}}, along with a partial documentation, also see Section 2.2 of \citet{blecicdobbsdixon2017}, and Chapter 5 of \citet{cubillos2017}.  \emph{Pyrat Bay} is a Python code for synthesizing emission and transmission spectra, and carrying out retrievals, which is already documented\footnote{\url{https://pcubillos.github.io/pyratbay/}}, but not yet publicly available.
The transmission spectrum part of the \emph{CHIMERA} code (e.g., \citealt{linewolf2013,linemarley2016}, but see \citealt{brogiline2018} for its most recent description) is available on github\footnote{\url{https://github.com/ExoCTK/chimera/}}, currently without documentation. \emph{HELIOS-R} is a retrieval code that was used for inferring properties of the HR~8799 planets via their emission spectra \citep{laviemendoca2017}. The code has been advertized to be publicly available at some point\footnote{\url{https://github.com/exoclime/HELIOS-R}}. \emph{PLATON} is a recently published Python retrieval package that calculates exoplanet transmission spectra \citep{zhangchachan2019}. It is available on github\footnote{\url{https://github.com/ideasrule/platon}}, and now also appears to allow for the calculation of emission spectra\footnote{\url{https://platon.readthedocs.io/en/latest/}}. The \emph{PLATON} code is designed to maximize computational speed, such that retrievals of transmission spectra can be carried out within minutes on a standard computer. This is done at the expense of the accuracy of the results: the opacity-sampling employed in \emph{PLATON} adds white noise, making their transit depths accurate to \rch{only} 100~\rch{parts per million (ppm)}. Due to the white-noise nature of the inaccuracies this should in first order only affect the width of their retrieved posterior distributions, and for retrievals with data of lower resolution than the intrinsic wavelength spacing of the code, opacity sampling inaccuracies are largely averaged out \citep{zhangchachan2019}. \rch{A detailed discussion of the effect of opacity sampling, and wavelength-averaged cross-sections, on the accuracy of spectral calculations, can be found in the recent study by \citet{garlandirwin2019}}.

\vspace{10mm}

In \ptrad, spectra can be calculated at low ($\lambda/\Delta \lambda=1000$) and high ($\lambda/\Delta \lambda=10^6$) resolution, using a correlated-k or line-by-line framework, respectively. The implementation of the radiative transfer, including the contribution functions, is described in Section \ref{sect:code_description}. The high-resolution part of \ptrad \ has already been used in \citet{mollieresnellen2018}. The opacity database, including line opacities, cloud opacities and gas quasi-continuum opacities, is described in Section \ref{sect:opacities}. At low or high resolution, spectra are calculated within a few seconds, scaling linearly with the wavelength coverage.\footnote{For the high resolution mode this describes the computational time over wavelength ranges typically covered by high resolution observations. We find, for example, 5 seconds for 90,000 wavelength points between 2.2 and 2.4 micron.} At low resolution, this speed is lower than the fastest publicly available radiative-transfer codes, for example \emph{PLATON} \citep{zhangchachan2019}, which takes a fraction of a second. However, as described above, the opacity sampling employed in their work adds white noise. \ptrad \ would be at least 16 times faster (the number of the correlated-k sub-bins is 16) if opacity sampling was employed. Because of the use of correlated-k, \ptrad \ low and high-resolution spectra agree excellently, see Section \ref{sect:code_verify}. In this section we also verify \ptrad \ by comparing to our self-consistent \emph{petitCODE} \citep{mollierevanboekel2015,mollierevanboekel2016}, also using models with condensate clouds. This verification means that \ptrad \ is also consistent with with the \emph{ATMO} \citep{tremblinamundsen2015} and \emph{Exo-REM} \citep{baudinobezard2015} codes, because \petit has been successfully benchmarked against these, see \citet{baudinomolliere2017}. \petit has also been used to carry out grid retrievals of self-luminous planets and brown dwarfs \citep{samlandmolliere2017}, the latter comparing well to the \emph{CHIMERA} results in \citet{lineteske2015}. After verification, we test retrievals of synthetic \emph{JWST} emission and transmission spectra with \ptrad, carried out on a small cluster (using 30 cores), see Section \ref{sect:ref_test}.  For computing quantities such as the planet-to-star flux ratio, or stellar heterogeneity effects \citep{rackhamapai2018}, \ptrad \ also includes a library of \emph{PHOENIX} \citep{husserwende-vonberg2013} and \emph{ATLAS9} \citep{kurucz1979,kurucz1992,kurucz1994} spectra, as described in \citet{vanboekel2012}. This library returns the stellar spectrum as a function of the stellar effective temperature.

\ptrad \ is available at \url{http://gitlab.com/mauricemolli/petitRADTRANS}, and its documentation can be found at \ptdoc.  Our Python retrieval implementation (see Section \ref{sect:ref_test}), using \ptrad \ and \emph{emcee} \citep{foreman-mackeyhogg2013}, can be found there as well. \ptrad \ makes use of the Numpy library \citep{numpy}.

\section{Code description}
\label{sect:code_description}
\ptrad \ consists of two resolution modes. The correlated-k mode (`\emph{c-k}') calculates spectra making use of the correlated-k approximation (see below), at a wavelength spacing of $\lambda/\Delta\lambda=1000$. The line-by-line mode (`\emph{lbl}') calculates the radiative transfer in a line-by-line fashion, that is directly in wavelength space. The line-by-line wavelength spacing is $\lambda/\Delta\lambda=10^6$.
\subsection{Emission spectra}
\subsubsection{Radiative transfer implementation}
For the calculation of the emission spectra we calculate the intensity along rays of different directions. For this we assume a plane-parallel atmosphere, that the atmosphere is in LTE, and neglect scattering, to speed up retrieval calculations. \rch{Neglecting scattering will mostly affect the blue part of the spectrum, and the near-infrared, if the atmospheres are cloudy, as we show in Section \ref{sect:comp_low_low_res}.} Scattering is included when calculating transmission spectra, see below. In a discrete-layer representation, the arising intensity in a plane-parallel atmosphere can be written as
\beq
\bar{I}^{\rm top} = \bar{B}(T_{\rm bot})\bar{\tr}^{\rm atmo}+\frac{1}{2}\sum_{i=0}^{N_{\rm L}-1}\left[\bar{B}(T^i)+\bar{B}(T^{i+1})\right]\left(\bar{\tr}^{i}-\bar{\tr}^{i+1}\right),
\label{equ:top_intensity}
\eeq
see, for example, Section 6.3.1 in \citet{molliere2017}. Here $N_{\rm L}$ is the number of atmospheric layers and the bar on top of a symbol (e.g., $\bar{B}$) denotes the wavelength average of that quantity, within the correlated-k spectral bin of interest. $B$ denotes the Planck function, and here we used that it is roughly constant across one of these bins, and replaced it with its mean value within the bin. $\tr$ denotes the atmospheric transmission from a given layer to the top of the atmosphere, \rch{and $\tr^{\rm atmo}$ is the transmission from the bottom to the top of the atmosphere.} Equation \ref{equ:top_intensity} is equivalent to Equation 13 in \citet{irwinteanby2008}. For the line-by-line calculations at high resolution, this equation is simply evaluated at every wavelength step.

Using the correlated-k assumption, and further making the standard assumption that the opacity distribution functions between given molecular species are uncorrelated \citep[e.g.,][]{lacis_oinas1991,fuliou1998}, we can write the transmission from a layer $i$, at pressure $P_i$, to the top of the atmosphere, at pressure $P=0$, as
\beq
\bar{\tr}^{i}=\prod_{j=1}^{N_{\rm spec}}\left[\sum_{l=0}^{N_g}{\rm exp}\left(-\int_0^{P_i}\frac{X_j\kappa_{lj}}{a\mu}dP\right)\Delta g_l\right] = \prod_{j=1}^{N_{\rm spec}} \bar{\tr}^{i}_j \ ,
\label{equ:trans_product}
\eeq
see, for example, Appendix A.5 in \citet{molliere2017}. Here $N_{\rm spec}$ is the number of species, $N_g$ the number of the Gaussian quadrature points in $g$-space (where $g$ is the coordinate of the cumulative opacity distribution function), $X$ the mass fraction of a given species in a given atmospheric layer, $a$ the gravitational acceleration within the atmosphere, and $\kappa$ a given species' opacity. The angle $\vartheta$ between the atmospheric normal and the ray is accounted for through the $1/\mu=1/{\rm cos}\vartheta$ in the exponential. In particular, from Equation \ref{equ:top_intensity} one sees that if emission is the only process of interest, and the atmospheric temperature structure is known, only the atmospheric transmissions and Planck functions need to be computed. The transmission of given layers $i$ to the top of the atmosphere themselves are the products of the transmissions $\bar{\tr}^{i}_j$ of the atmosphere's individual opacity species $j$. This means that the numerically expensive computation of a combined correlated-k opacity table is not necessary. This allows for a fast correlated-k application in retrieval calculations.

For the line-by-line calculations at high resolution we do not employ this ``product of transmissions'' method. Rather, we calculate the total opacity by simply adding opacities of all opacity species in every spectral bin, and then calculate the transmission of the atmosphere. This conserves the wavelength correlation between the opacities of all species, and is hence not subject to the classical correlated-k assumption of uncorrelated opacity distributions.

We solve the radiative \rch{transfer} on a three-point Gaussian grid for the angle $\mu={\rm cos}\vartheta$ between rays and the atmospheric normal, for which we found only small differences when comparing to the 20-point Gaussian grid used in \emph{petitCODE}, see Section \ref{sect:code_verify}. The flux is then calculated as
\beq
F = 2\pi \sum_{k=1}^{N_\mu} \mu_k \bar{I}^{\rm top}(\mu_k)\Delta \mu_k,
\label{equ:num_flux}
\eeq
which follows from $F = 2\pi\int_0^1 \mu I(\mu)d\mu$.

\subsubsection{Contribution function implementation}
The emission contribution function quantifies the relative importance (contribution) of the emission in a given layer to the total atmospheric flux.
Using equations \ref{equ:top_intensity} and \ref{equ:num_flux}, one can see that the relative contribution of a given layer, at spatial coordinate $i+1/2$, to the total flux is
\beq
C_{\rm em}^{i+1/2} = \frac{\sum_{k=1}^{N_\mu} c^{i+1/2}(\mu_k)\mu_k\Delta\mu_k}{\sum_{l=1}^{N_\mu}\left[2\bar{B}(T_{\rm bot})\bar{\mathcal{T}}^{\rm atmo}(\mu_l)+\sum_{j=1}^{N_{\rm L}-1}c^{j+1/2}(\mu_l)\right]\mu_l\Delta\mu_l},
\label{equ:emis_contr}
\eeq
where
\beq
c^{i+1/2}(\mu_k) = \left[\bar{B}(T^i)+\bar{B}(T^{i+1})\right]\left[\bar{\mathcal{T}}^{i}(\mu_k)-\bar{\mathcal{T}}^{i+1}(\mu_k)\right].
\eeq
These equations are evaluated in \ptrad \ when the emission contribution function is calculated.

\begin{figure}[t!]
\centering
\includegraphics[width=0.5\textwidth]{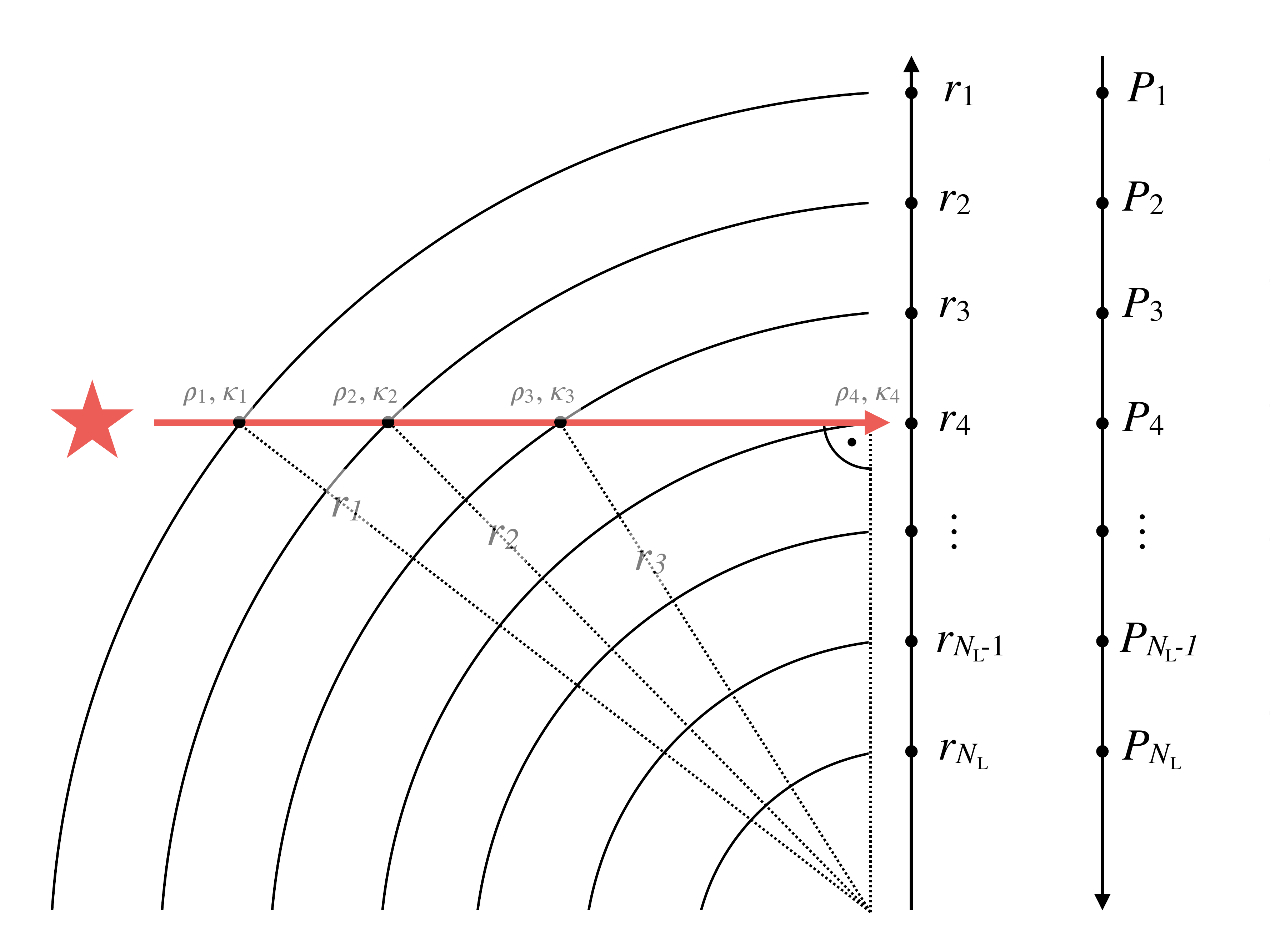}
\caption{Geometry of the transmission problem using the same notation as in Section \ref{sect:transm_spec}. The red line shows a grazing light ray as it passes through the planetary atmosphere.}
\label{fig:transm_geo}
\end{figure}

\subsection{Transmission spectra}
\label{sect:transm_spec}
\subsubsection{Radiative transfer implementation}
For the calculation of transmission spectra, the input 1-d temperature and abundance structure is assumed to describe the atmospheric vertical structure at all longitudes and latitudes. The pressure-temperature structure is converted into a radius-temperature structure assuming hydrostatic equilibrium, mapping the pressure $P_i$ of a layer $i$ to the radius $r_i$. For this, a reference radius and planetary gravity at an arbitrary pressure must be specified. The optical depth of a ray of light, grazing the atmosphere at impact parameter $r_i$ above the planetary center, can then be written as
\beq
\tau_i = \sum_{j=1}^{i-1}\left(\kappa_j\rho_j+\kappa_{j+1}\rho_{j+1}\right)\left(\sqrt{r_j^2-r_i^2}-\sqrt{r_{j+1}^2-r_i^2}\right),
\eeq
where $\rho_j$ is the mass density of the atmospheric layer at radius $r_j$. The geometry of the transmission problem is sketched in Figure~\ref{fig:transm_geo}. \rch{The factor $1/2$ of the integrand approximation $(\kappa_j\rho_j+\kappa_{j+1}\rho_{j+1})/2$ cancels with the factor $2$ in front of the integral, for grazing rays approaching and receding from the point of closest approach to the planet's center.}

The effective area of the planet is calculated as
\beq
A = 2\pi \int_0^{R_{\rm pl}} r \left(1-\tr\right)dr,
\label{equ:transmission_area}
\eeq
where $\tr=e^{-\tau}$ again denotes the transmission of light through the atmosphere, this time in grazing geometry. As before it is possible to write the transmission as the product of the individual species' transmissions, greatly speeding up the correlated-k treatment in the process.

Again, for the line-by-line calculations at high resolution we do not employ this ``product of transmissions'' method. Rather, we calculate the total opacity by simply adding opacities of all opacity species in every spectral bin, and then calculate the transmission of the atmosphere.

In \ptrad , Equation \ref{equ:transmission_area} is solved numerically. This is done by starting from the bottom of the atmosphere, assuming that the planet is opaque at all wavelengths at the lowest altitude.

\subsubsection{Contribution function implementation}

Transmission contribution functions can be defined in multiple ways, for example by determining the pressure for which a certain optical depth is reached, in transit geometry. We suggest and use different method here, which quantitatively measures the importance of a given layer when forming the transmission spectrum.

For a given layer $i$, the transmission contribution at a given wavelength is calculated using
\beq
C_{\rm tr}^{i} = \frac{R_{\rm nom}^2-R^2(\kappa_i=0)}{\sum_{j=1}^{N_{\rm L}}\left[R_{\rm nom}^2-R^2(\kappa_j=0)\right]} \ ,
\label{equ:trans_contr}
\eeq
where $R_{\rm nom}$ is the planet's nominal transmission radius at a given wavelength, and $R(\kappa_i=0)$ is the transmission radius one obtains when setting the opacity in the $i$th layer to zero, and recalculating the transmission spectrum. Squared radii are used here, because transmission spectra measure the flux decrease of the star as it is transited by its planet, which is proportional to the planet's area.

Calculating the transmission function in the way presented here has the advantage that only those layers which can actually change the transmission radius are assigned a large contribution. High altitude layers will not contribute strongly, because they are optically thin, whereas low altitude layers will be negligible because they are hidden from view by the overlying layers.

\section{Opacities}
\label{sect:opacities}

Below we describe the references for the various opacity sources available within \ptrad, which encompass molecular and atomic line opacities, cloud opacities (absorption and scattering), Rayleigh scattering cross-sections, and quasi-continuum opacities (collision induced absorption, H$^-$ \rch{bound-free} (b-f) and \rch{free-free} (f-f) absorption). The opacities are published together with the code.

\begin{table*}[t!]
\centering
{ \footnotesize
\begin{tabular}{cccccc}
\hline \hline
Opacity source & Spectral range [$\mu$m] & Line list & Partition function & Pressure broadening & Line Cutoff \\ \hline
CH$_4$ & 0.83 $< \lambda$ & YT14 & F03 & Eq. (15), SB07 & HB02 \\
C$_2$H$_2$ &  1 $< \lambda < $ 16.5 & \emph{HITRAN} & F03 & $\gamma_{\rm air}$ & HB02 \\
CO  &  1.18 $< \lambda$ & \emph{HITEMP} & F03 & $\gamma_{\rm air}$ & HB02 \\
CO & 0.112 $< \lambda < $ 0.43 & K93 & F03 & Eq. (15), SB07 & HB02 \\
CO$_2$ & 1 $< \lambda < $ 38.76 & \emph{HITEMP} & F03 & $\gamma_{\rm air}$ & BG69 \\
H$_2$S & 0.88 $< \lambda$ & \emph{HITRAN} & F03 & $\gamma_{\rm air}$ & HB02 \\
H$_2$ & 0.28 $< \lambda$ & \emph{HITRAN} & F03 & $\gamma_{\rm air}$ & HB02 \\
H$_2$ & 0.08 $< \lambda < $ 0.18 & K93 & F03 & Eq. (15), SB07 & HB02 \\
HCN & 2.92 $< \lambda$ & HT06, BS14 & F03 & Eq. (15), SB07 & HB02 \\
H$_2$O & 0.33 $< \lambda$ & \emph{HITEMP} & F03 & $\gamma_{\rm air}$ & HB02 \\
K & 0.05 $< \lambda$ & PK95 & ST84 & N. Allard$^{\rm (a)}$, SH96 & HB02 \\
Na & 0.1 $< \lambda$ & PK95 & ST84 & N. Allard, SH96 &  HB02 \\
NH$_3$ & 0.8 $< \lambda$ & YB11 & SH14 & Eq. (15), SB07 & HB02 \\
O$_3$ & 1.43 $< \lambda$ & \emph{HITRAN} & F03 & $\gamma_{\rm air}$ & HB02 \\
OH & 0.52 $< \lambda$ & \emph{HITEMP} & F03 & $\gamma_{\rm air}$ & HB02 \\
PH$_3$ & 1 $< \lambda$ & SA15 & F03 & Eq. (15), SB07 & HB02 \\
TiO & 0.32 $< \lambda$ & B. Plez$^{\rm (a)}$ & U. J{\o}rgensen$^{\rm (b)}$ & Eq. (15), SB07 & HB02 \\
VO & 0.36 $< \lambda <$ 2.6 & B. Plez & B. Plez & Eq. (15), SB07 & HB02 \\
SiO & 1.6 $< \lambda$ & BY13 & BY13 & Eq. (15), SB07 & HB02 \\
FeH & 0.67 $< \lambda$ & WR10 & WR10 & Eq. (15), SB07 & HB02 \\
\hline
\end{tabular}
}
\caption{References for the atomic and molecular opacities available for use in \ptrad. Reference codes: \emph{HITEMP} \citet{rothman2010}, \emph{HITRAN}: \citet{rothman2013}, SB07: \citet{sharpburrows2007}, F03: \citet{fischer2003}, ST84: \citet{sauval_tatum1984}, K93: \citet{kurucz1993}, YT14: \citet{yurchenko2014}, YB11: \citet{yurchenkobarber2011}, SH96: \citet{schweitzer1996}, SH14: \citet{sousa-silvahseketh14}, SA15: \citet{sousa-silvaal-refaie14}, HT06: \citet{harris2006}, BS14: \citet{barber2014}, PK95: \citet{piskunov1995}, HB02: \citet{hartmannboulet2002}, BG69: \citet{burchgryvnak1969}, BY13: \citet{bartonyurchenko2013}, WR10: \citet{wendereiners2010}, (a): priv. comm. (b): partition function retrievable from \url{http://www.astro.ku.dk/~uffegj/scan/scan_tio.pdf}. For all molecules, except for CO and TiO, only the main isotopologue opacities are used in the low resolution mode. At high resolution also secondary isotopologues are available for some species, see the documentation at \ptdoc \ for an up-to-date list. \rch{The documentation website also contains a short tutorial of the \emph{ExoCross} code \citep{yurchenkoalrefaie2018}, and how to convert its resulting \emph{Exomol} opacity calculations for use in \ptrad.} \rcht{This allows the \ptrad \ user to calculate line opacities of additional species, for the low and high resolution modes.}
}
\label{tab:line_opa_sources}

\centering
{ \footnotesize
\begin{tabular}{lc|lc|lc}
\hline \hline
Quasi-continuum &  Reference & Clouds & Reference & Rayleigh & Reference\\ \hline
H$_2$--H$_2$ CIA & BR, RG12 & Al$_2$O$_3$ & \citet{koikekaito1995} & H$_2$ & \citet{dalgarnowilliams1962}  \\
H$_2$--He CIA & BR, RG12 & H$_2$O (ice) & \citet{smithrobinson1994} & He & \citet{chandalgarno1965} \\
H$^-$ bound-free & \citet{gray08}  & Fe & \citet{henningstognienko1996} & CO & \citet{sneepubachs2005}  \\
H$^-$ free-free & \citet{gray08} & KCl & \citet{palik2012} & CO$_2$ & \citet{sneepubachs2005} \\
& & MgAl$_2$O$_4$ & \citet{palik2012} & CH$_4$ & \citet{sneepubachs2005} \\
& & MgSiO$_3$ & \citet{scottduley1996}, & H$_2$O & \citet{harveygallagher1998} \\
& & & \citet{jaegermolster1998} & & \\
& & Mg$_2$SiO$_4$ & \citet{servoinpiriou1973} & O$_2$ & \citet{thalmanzarzana14,thalmanzarzana17} \\
& & Na$_2$S & \citet{morleyfortney2012} & N$_2$ & \citet{thalmanzarzana14,thalmanzarzana17} \\
& & SiC & \citet{pegourie1988} & & \\ \hline
\end{tabular}
}
\caption{Quasi-continuum, Rayleigh and cloud opacities available in \ptrad. References BR stand for: \citet{borysow1988,borysow1989a,borysow1989b,borysow2001,borysow2002}, while RG12 stands for \citet{richardgordon2012}, and the references therein.}
\label{tab:opa_sources_cont}

\end{table*}

\subsection{Line opacities}
\label{sect:line_opas}
\ptrad \ allows to include the line opacities of Na, K, \ce{CH4}, \ce{C2H2}, CO, \ce{CO2}, \ce{H2S}, \ce{H2}, HCN, \ce{H2O}, \ce{NH3}, \ce{O3}, OH, \ce{PH3}, TiO, VO, SiO and FeH. High temperature linelists are used when available, see Table \ref{tab:line_opa_sources} for the references. For all molecules, except for CO and TiO, only the main isotopologue opacities are used in the low-resolution mode. For CO and TiO all isotopologues where included, at telluric occurrence rates, because here we found that they add a signigcant amount of opacity already at low resolution. In the high-resolution mode, also secondary isotopologues are available, for example for \ce{H2O}, \ce{CH4} and \ce{CO}. See the code documentation website at \ptdoc \ for the up-to-date list of high-resolution species.

\rch{The documentation website also contains a short tutorial of the \emph{ExoCross} code \citep{yurchenkoalrefaie2018}, and how to convert its resulting \emph{Exomol} opacity calculations for use in \ptrad.} \rcht{This allows the \ptrad \ user to calculate line opacities of additional species, for the low and high resolution modes.}

Pressure broadening is included either by using the air broadening coefficients of \emph{HITRAN}/\emph{HITEMP} \citep{rothman2010,rothman2013}, or Equation 15 of \citet{sharpburrows2007}. Using the air broadening coefficients $\gamma_{\rm air}$ may sound like a stretch for the often \ce{H2}/He-dominated exoplanet atmospheres. However, \citet{gharibline2018} have shown that the ratio between air and \ce{H2}/He broadening ranges from values of 1 to 2 (see their Table 1). Nonetheless, we plan to include \ce{H2}/He broadening in future versions of the opacity database. \citet{gharibline2018} also showed that care must be taken if the atmospheres are strongly enriched: self-broadening of water then becomes much stronger than that of \ce{H2}/He, or air.

For all molecules, except for \ce{CO2}, a line cutoff is included as an exponential line wing decrease, as measured for \ce{CH4} when broadened by \ce{H2} \citep{hartmannboulet2002}.
For \ce{CO2}, the line wing decrease is modeled using a fit to the CO$_2$ measurements by \citet{burchgryvnak1969}. This fit was obtained from Bruno B\'ezard (private communication).

The line opacities are calculated from 110~nm to 250~$\mu$m, 80-3000~K, and $10^{-6}$-$10^3$ bar. \rch{In the high-resolution mode ($\lambda/\Delta \lambda=10^6$), the opacities only range from 0.3 to 28 micron.} The pressure and temperature points are spaced equidistantly in log-space on a $10\times13$ point grid. If a $P$-$T$ structure is defined that goes outside of the $P$ and $T$ range mentioned here, \ptrad \ will use the opacities at the boundary of the grid. \rch{Given this upper temperature boundary, \ptrad \ can calculate the fluxes of terrestrial planets, super-Earths, Neptunes, and hot Jupiters}\rcht{, if their atmospheres are hydrogen-dominated}\rch{. Higher temperature points and species will be added in the future, increasing the temperature resolution and range of the existing grid, and allowing to also model the class of ultra hot-Jupiters,} \rcht{with equilibrium temperatures} \rch{$T_{\rm eq} \gtrsim 2000$~K \citep[see, e.g.,][]{arcangelidesert2018,lothringerbarman2018,lothringerbarman2019}. Thus, for now, the results of spectral calculations and retrievals of {\it ultra} hot Jupiters must be treated very carefully. This is because the spectrally active regions can reach temperatures larger than 3000 K for planets of equilibrium temperatures of about 2000~K and above.}

\begin{figure*}[t!]
\centering
\includegraphics[width=1.\textwidth]{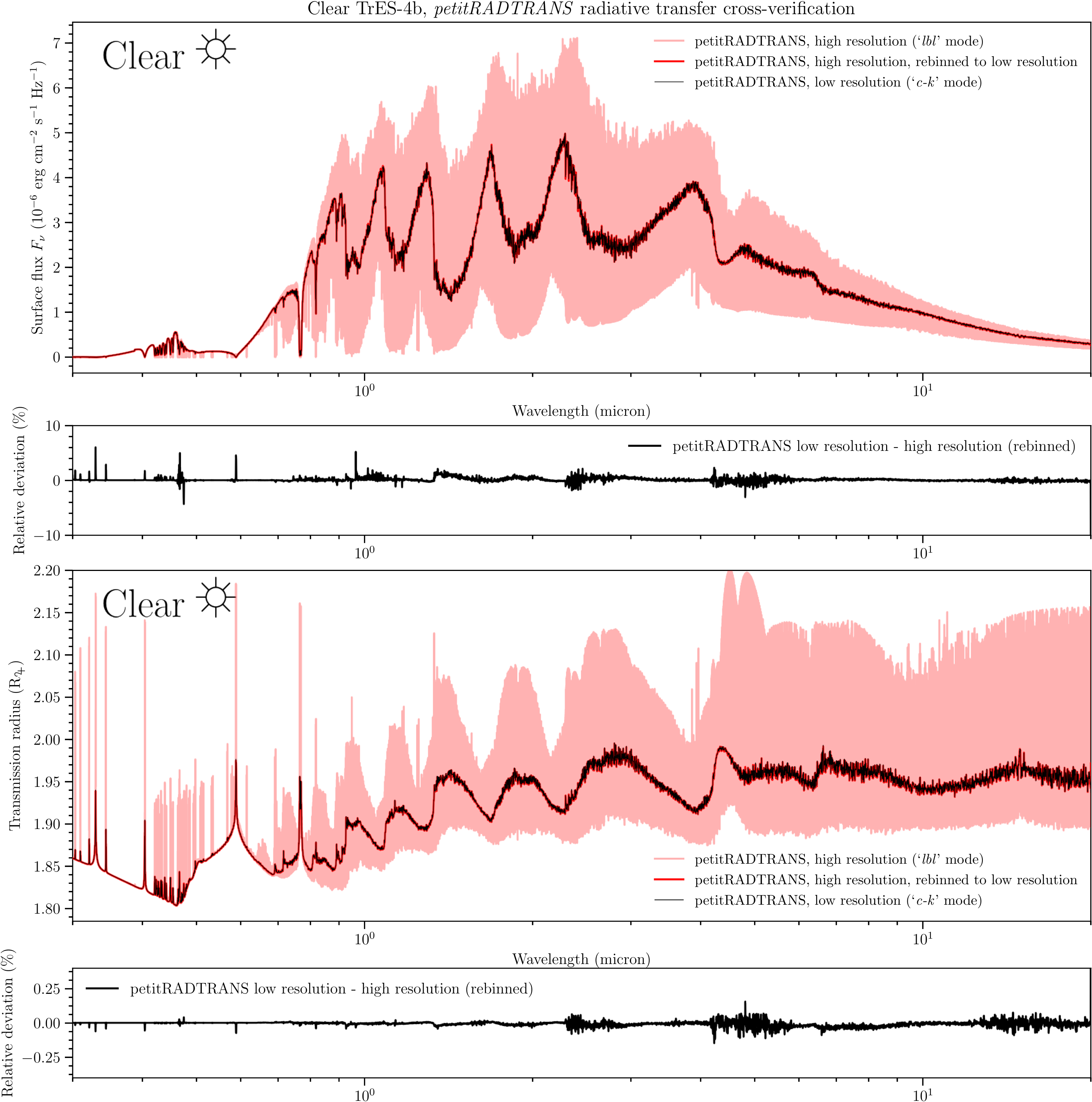}
\caption{Radiative transfer verification of \ptrad. The {\it uppermost panel} shows the comparison between the emission spectra of a clear TrES-4b-like atmosphere, calculated with \ptrad \ in high-resolution (`\emph{lbl}`) mode (light red), and in low-resolution (`\emph{c-k}`) mode (black, thin line). The red line denotes the re-binned high-resolution spectrum. The panel below shows the residuals between the two modes. The {\it second lowest panel} shows the comparison between the transmission spectra of the same atmosphere, for both modes. The \emph{lowest panel} shows the residuals between the two modes, for the transmission spectrum.}
\label{fig:pt_rad_ver_lohi}
\end{figure*}

At low resolution, a wavelength binning of $\lambda/\Delta \lambda=1000$ is used. Within the corresponding wavelength bins, a 16-point Gaussian quadrature grid is used for the $g$ coordinate of the cumulative opacity distribution function. This $g$-grid is separated into an eight-point grid ranging from $g=0$ to $g=0.9$ and a second eight-point grid ranging from $g=0.9$ to $g=1$. At high resolution the wavelength binning of the line-by-line opacities is $\lambda/\Delta \lambda=10^6$.

\subsection{Rayleigh Scattering}

The Rayleigh cross-sections of the following atoms and molecules can be included in \ptrad \ calculations: \ce{H2}, He, CO, \ce{CO2}, \ce{CH4}, \ce{H2O}, \ce{O2} and \ce{N2}. See the right column of Table \ref{tab:opa_sources_cont} for the references of the respective cross-sections.

\subsection{Cloud opacities}
\label{sect:cloud_opas}
\ptrad \ allows for including clouds in four different ways:
\begin{enumerate}
    \item By defining a \rch{gray} cloud deck pressure, below which the atmosphere cannot be probed (effectively we add an absorption opacity $\kappa=10^{99}$ cm$^2$/g \rch{at all wavelengths}, there and below). This can also be used to implement a maximum visible pressure in transmission spectra, due to atmospheric diffraction, see, for example, Equation 15 in \citet{robinsonfortney2017}.
    \item By scaling the value of the Rayleigh scattering cross-section of the gas
    \beq
    \kappa = f \cdot \kappa_{\rm Rayleigh}(\lambda),
    \eeq
    where $f$ is the scaling factor and $\kappa_{\rm Rayleigh}$ is the Rayleigh opacity of the atmospheric gas. In this case the Rayleigh scattering opacity will hence be $f$ times larger. 
    \item By introducing a scattering cross-section with two free parameters $\kappa_0$ and $\gamma$,
    \beq
    \kappa = \kappa_0\left(\frac{\lambda}{\lambda_0}\right)^\gamma,
    \eeq
    where $\kappa_0$ is the opacity at 0.35~$\mu$m, in units of $\rm cm^2/g$. This is equivalent to the parametrization used in, for example, \citet{macdonaldmadhusudhan2017}. 
    \item By including solid \ce{Al2O3}, \ce{H2O}, Fe, KCl, \ce{MgAl2O4}, \ce{MgSiO3}, \ce{Mg2SiO4} or \ce{Na2S} clouds, for which the optical constants given in the middle column of Table \ref{tab:opa_sources_cont} are used. For this mode, the altitude-dependent cloud mass fraction, mean radius of the particles, and the width of the log-normal particle size distribution, need to be specified. The particle radius can also be calculated by specifying the atmospheric eddy diffusion coefficient $K_{zz}$ and settling parameter $f_{\rm sed}$, as described and defined in \citet{ackermanmarley2001}.
\end{enumerate}
The four cloud modes described above can be used independently, but also in combination. For emission spectra only the absorption component of the cloud options listed above will be used, if present \rch{(that is, options 1 and 4)}. It is also possible to test for the importance of scattering by adding the scattering opacity as an absorption opacity.

The cloud opacities of the species listed in Table \ref{tab:opa_sources_cont} are calculated assuming either homogeneous and spherical, or irregularly shaped cloud particles.  Comparing spectra using these different cloud opacity treatments could allow for the distinction between particle shapes. This holds for particles of small enough grain sizes ($\lesssim 1 \ \mu {\rm m}$), for which the resonance features of the cloud species are most clearly visible \citep{minhovenier2005}.

The opacity of the irregularly shaped cloud particles is approximated by taking the opacities obtained for a Distribution of Hollow Spheres (DHS). For spherical and DHS cloud particles, the opacities were calculated with the code of \citet{minhovenier2005}, which also uses software reported in \citet{toonackerman1981}. Spherical particle cross-section are obtained from Mie theory, while an extended Mie formulation is assumed in the DHS case. For DHS, a porosity of $P=0.25$ \citep[as in][]{woitkemin2016} and an irregularity parameter $f_{\rm max}=0.8$, as defined in \citet{minhovenier2005}, are used.

The opacities were calculated for particles of sizes between 1 nm and 10 cm. If the \ptrad \ user requests opacities for particles with sizes outside this range, the returned opacity for the corresponding cloud species will be zero.\footnote{It is also possible to request cloud opacities for a log-normal size distribution. In this case all size ranges outside of the 1 nm to 10 cm interval will not contribute to the total opacity.}

\subsection{Other continuum opacities}
\ptrad \ allows to include \ce{H2}-\ce{H2} and \ce{H2}-He Collision Induced Absorption (CIA) cross-sections, as well as the bound-free and free-free absorption of \ce{H-}. See the left column of Table \ref{tab:opa_sources_cont} for the references.

\section{Code verification}
\label{sect:code_verify}

In Section \ref{sect:comp_high_low_res} we compare the low-resolution (`\emph{c-k}') mode to the high-resolution (`\emph{lbl}') mode of \ptrad. We hence verify our correlated-k (at low resolution) and line-by-line (at high resolution) implementation, because binned high-resolution spectra should be equal to the correlated-k results at lower resolution.

In Section \ref{sect:comp_low_low_res} we compare the low resolution (`\emph{c-k}') mode of \ptrad \ with the radiative transfer calculations obtained from \emph{petitCODE}. \petit is a 1-d, self-consistent atmospheric code, solving for the atmospheric structure in radiative-convective and chemical equilibrium \citep{mollierevanboekel2015,mollierevanboekel2016}. \petit radiative transfer includes scattering, and makes use of the correlated-k assumption.
In \citet{baudinomolliere2017}, it has been successfully benchmarked with the \emph{ATMO} \citep{tremblinamundsen2015} and \emph{Exo-REM} \citep{baudinobezard2015} codes. Moreover, \petit results have themselves been verified by comparing to line-by-line, high-resolution ($\lambda/\Delta\lambda=10^6$) calculations, which where binned to \emph{petitCODE}'s wavelength sampling \citep{mollierevanboekel2015}.
See Table~\ref{tab:code_comp} for a comparison between the modeling approaches in \ptrad \ and \emph{petitCODE}.

For the comparisons presented here, the input atmospheric temperature and abundance structures where obtained from a self-consistent \petit solution, assuming $T_* = 6295$~K, $R_* = 1.831$~R$_\odot$, $d=0.0516$~au, $T_{\rm int}=100$~K, $M_{\rm P} = 0.494$~$\mj$, $R_{\rm P} = 1.838$~$\rj$. We assumed a planetary metallicity of [Fe/H] = 1.1 and a solar C/O ratio (0.55). We assumed a dayside redistribution of the stellar flux. These parameters where taken from our TrES-4b model in \citet{mollierevanboekel2016}.

\subsection{Comparing the high (`\emph{lbl}') and low (`\emph{c-k}') resolution modes}
\label{sect:comp_high_low_res}

Figure \ref{fig:pt_rad_ver_lohi} shows the comparison between the high (`\emph{lbl}') and low (`\emph{c-k}') resolution modes of \ptrad, for the clear TrES-4b-like atmosphere. One can see that both emission and transmission spectra agree very well, with the errors scattering around zero, and in the low, single-digit percentage range, as usually found for correlated-k implementations \citep[see, e.g.,][]{lacis_oinas1991,fuliou1998,mollierevanboekel2015}.

\subsection{Comparing \ptrad \ to \emph{petitCODE}}
\label{sect:comp_low_low_res}

\begin{figure*}[t!]
\centering
\includegraphics[width=1.\textwidth]{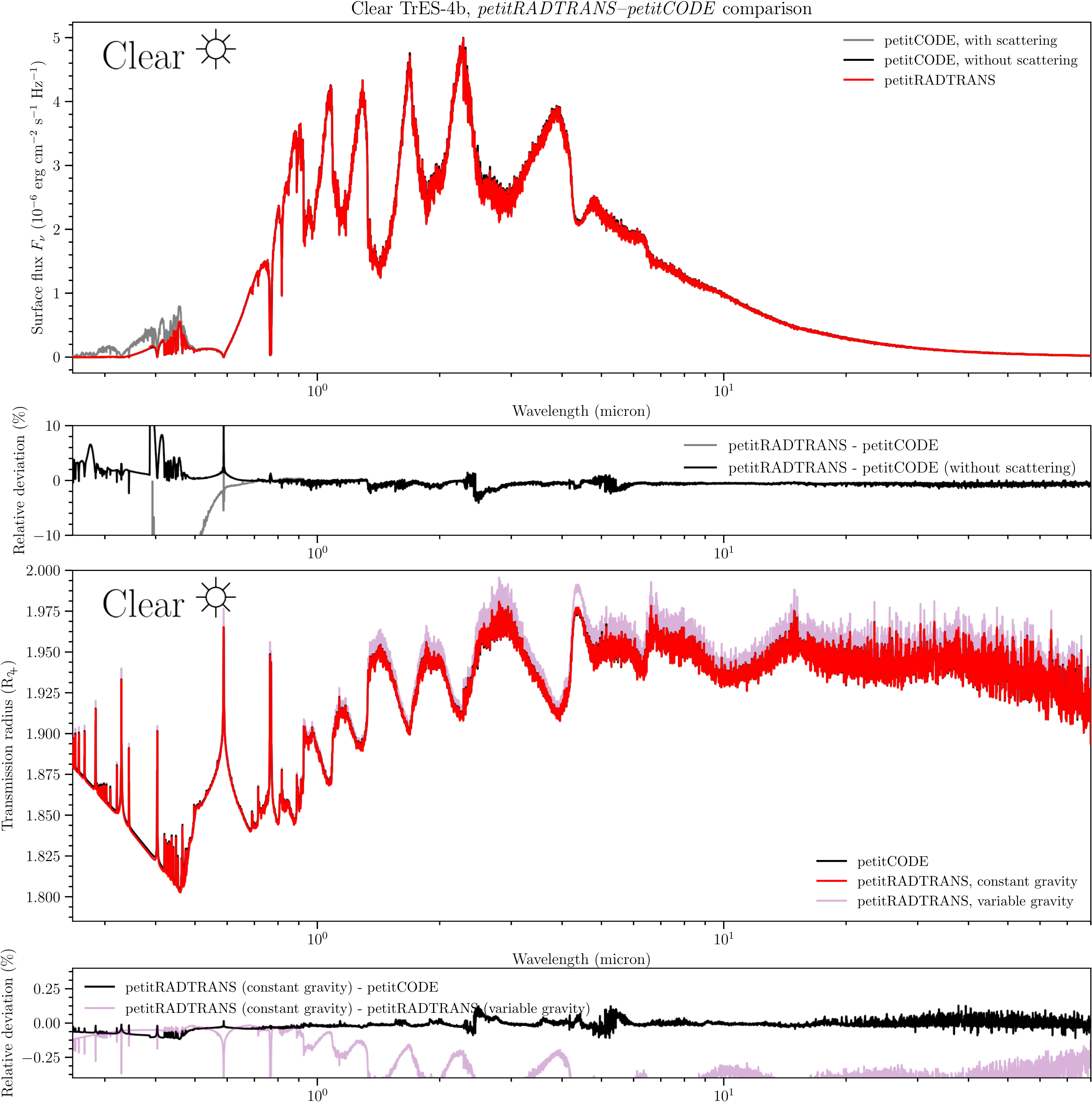}
\caption{Radiative transfer verification of \ptrad. The {\it uppermost panel} shows the comparison between the emission spectra of a clear TrES-4b-like atmosphere, calculated with \ptrad \ (red), \petit without scattering (black) and \petit with scattering (gray). The agreement between \petit (with scattering) and \ptrad \ breaks down for wavelengths below 0.5 $\mu$m, because Rayleigh scattering becomes important (\ptrad \ does not include scattering for emission spectra). The {\it second lowest panel} shows the comparison between the transmission spectra of the same atmosphere, calculated with \ptrad \ when assuming a constant surface gravity (red), when assuming a variable gravity (purple), and with \emph{petitCODE}, which assumes a constant gravity (black). Here, the constant gravity cases agree well also in the short wavelength range, because scattering is included for transmission spectra in \ptrad. The comparison between the constant and variable gravity cases of \ptrad \ highlights the effect of varying gravity on the resulting spectra. The panels below the emission and transmission spectra show the relative deviation of the \ptrad \ calculations, which are shown as red lines in the spectral plots, to the other cases.}
\label{fig:pt_rad_ver_clear}
\end{figure*}

\begin{figure*}[t!]
\centering
\includegraphics[width=1.\textwidth]{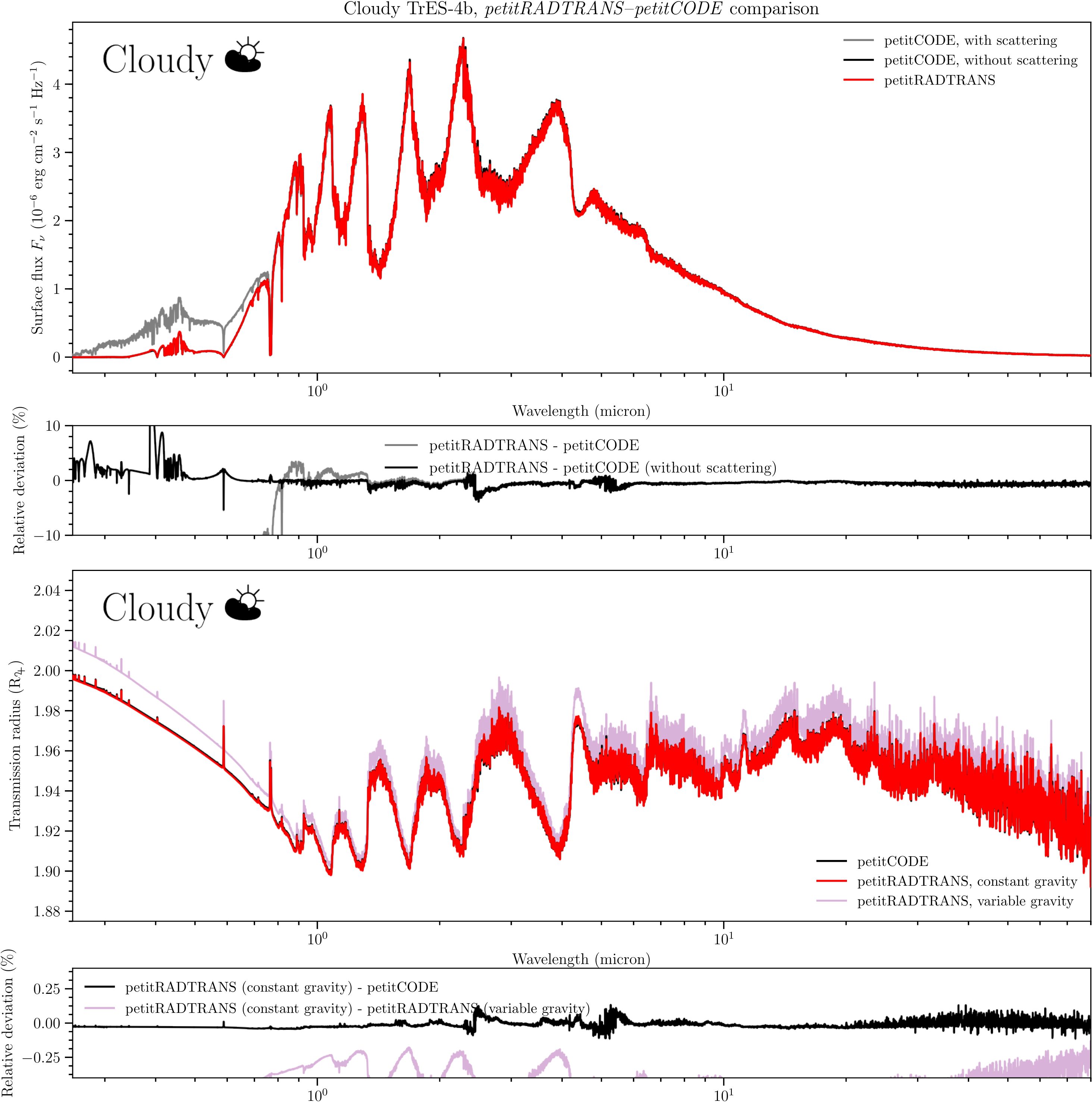}
\caption{Same as Figure \ref{fig:pt_rad_ver_clear}, but for the cloudy case.}
\label{fig:pt_rad_ver_cloudy}
\end{figure*}

\subsubsection{Clear atmospheres}

The upper panel of Figure \ref{fig:pt_rad_ver_clear} shows a comparison between \petit and \ptrad \ emission spectra. The \petit emission spectra are plotted with and without scattering. The \ptrad \ spectra do not include scattering. The differences between \ptrad \ and \petit are small (see the second panel from the top), usually on the order of 1~\% or less. There are regions where deviations are larger, up to 4~\%, especially at around 2.3 and 4.5 microns, that is at the location of the fundamental and first overtone band of CO. These differences vanish in calculations neglecting CO in both \petit and \ptrad. We checked multiple times and are certain that we use the same CO opacities in both codes. Differences in the numerical radiative transfer treatment between of \ptrad \ and \emph{petitCODE} may be responsible for these small differences, but prove difficult to identify. We note that \petit uses a 20 point Gaussian quadrature grid for the correlated-k $g$ coordinate \citep{mollierevanboekel2015}, whereas \ptrad \ uses 16 points (see Section \ref{sect:line_opas}). However, we have verified in Section \ref{sect:comp_high_low_res} that the correlated-k implementation of \ptrad \ reproduces the line-by-line results very accurately. Moreover, the differences in solving the radiative transfer equation for emission spectra are not a likely candidate (Feautrier method in \emph{petitCODE}, Equation \ref{equ:top_intensity} in \ptrad), because the same wavelengths show small systematic differences for the transmission spectra, see discussion below. \rch{These differences could point to a somewhat smaller accuracy of the \emph{petitCODE} opacity treatment, when compared with \ptrad. In \petit the opacities are combined to yield the correlated-k opacity table for the full gas mixture, which is needed for calculating the atmospheric temperature structure, and scattering source function. For this combination we assume that the individual species' opacities are not correlated in wavelength space. The same mathematical assumption pertains to the ``multiplication of transmissions'' treatment of \ptrad, but it may result in a different behavior, in terms of numerical errors.}

Hot Jupiters typically have planet-to-star contrasts between $10^{-3}$ and $10^{-4}$. Flux differences of up to 4~\%, as we obtain for the \ptrad--\petit comparison here, hence lead to flux differences between 4 to 40 ppm.

Because the model considered here is cloud-free, scattering does not play an important role for the majority of the spectrum. Only for $\lambda<0.5$~$\mu$m significant differences between \petit \ (with scattering, here leading to reflection of the stellar light) and \ptrad \ (without scattering) start to appear \rch{(up to $\sim$90~\% relative deviation)}. We note that not much flux is escaping from the planetary atmosphere at such small wavelengths. Hence we expect that \ptrad \ can calculate emission spectra of cloud-free hot Jupiters within the wavelength range of the upcoming \emph{JWST} \citep{beichmanbenneke2014}.

There also appear to be some larger differences in the wings of the \rch{alkali lines of secondary importance (at $\sim$0.4~micron)}, which do not occur for the transmission spectrum implementation, see below. Here we again tested for various effects that could likely explain this difference. We could rule out, for example, that it is the smaller number of ray angles used for calculating the radiation field (20 in \petit compared to 3 in \ptrad), because using as many angles as in \petit did not change the results.

In the two lower panels of Figure \ref{fig:pt_rad_ver_clear} the comparison of the transmission spectra of \ptrad \ and \petit is shown. In \ptrad \ transmission spectra the surface gravity is varying as a function of altitude, with an $r^{-2}$ dependence, where $r$ is the planet's radial coordinate. In contrast, \petit is a code which solves for the atmospheric structure assuming a plane-parallel atmosphere, using the pressure as its vertical coordinate. For this, the variation of the surface gravity is neglected, also during the calculation of the \petit transmission spectra. We hence show two different \ptrad \  calculations when comparing \petit and \ptrad \ spectra: one with varying surface gravity (nominal) and one where the surface gravity is constant. The reference pressure for the transmission spectra was chosen to be $P_0 = 0.01$~bar for \petit and \ptrad, such that $r(P_0) = R_{\rm P}$ and $g(r(P_0))=-GM_{\rm P}/R_{\rm P}^2$.

As can be seen, the agreement between \petit and \ptrad \ (with constant gravity) is of similar quality as in the emission case. The optical part ($\lambda < 0.5$~$\mu$m) is reproduced better than in the emission case, because photon extinction due to scattering is included in the \ptrad \ calculations for transmission spectra.

The difference of including the effect of varying gravity is visible as well, when comparing the constant and variable gravity calculations of \ptrad: in wavelength regions where the opacity is large, smaller pressures than the reference pressure are probed. Here the deviation from the constant surface gravity assumption is most visible.

\subsubsection{Cloudy atmospheres}
\label{sect:cloudy_atmos}

In this section we discuss the comparison between \ptrad \ and \emph{petitCODE}, for cloudy atmospheres. We used the same atmospheric setup for \petit as for the clear case, but self-consistently included clouds of \ce{Na2S}, \ce{KCl}, \ce{Mg2SiO4},  and \ce{MgAl2O4} in the atmospheric structure iteration of \emph{petitCODE}. For this we used Cloud Model 6, as defined in Table 2 of \citet{mollierevanboekel2016}, which corresponds to clouds of single particles size, with particle radii of 0.08~$\mu$m. \rch{This particle size was chosen because it highlights how small silicate particles can lead both to strong (Rayleigh) scattering in the optical, as well as silicate resonance features in the mid-infradred.} The cloud location in the atmosphere is coupled to equilibrium chemistry, that is the clouds can only exist in regions where the condensates do not evaporate. The cloud mass fraction is equal to the condensates' mass fraction in chemical equilibrium, but is capped at a maximum value of $3\times 10^{-4}\cdot Z_{\rm P}$, where $Z_{\rm P}$ is the planet's metal mass fraction. This upper value can thus be thought of as effectively parametrizing the particle settling strength. \rch{The value chosen here represents and intermediate choice for the cloud settling strength, also see Table 2 in \citet{mollierevanboekel2016}.} For the calculations shown here, irregularly shaped, crystalline particles were assumed, for which we made use of the DHS opacities (see Section \ref{sect:cloud_opas}).

The corresponding spectra for the cloudy case are shown in Figure \ref{fig:pt_rad_ver_cloudy}. For the emission case we again see a good agreement between \petit (including scattering) and \ptrad, with the difference that scattering (mainly leading to the reflection of stellar light) now already becomes important at larger wavelengths (for $\lambda<0.8$~$\mu$m), due to the added haze-like scattering of the clouds. It is noteworthy that the agreement at longer wavelengths is very good, although the transmission spectrum of the planet shows the significant impact of clouds (cf. clear transmission spectrum in Figure \ref{fig:pt_rad_ver_clear}), also the 10-micron feature of \ce{Mg2SiO4} \citep[e.g.,][]{wakefordsing2015,mollierevanboekel2016} is visible. This difference in the visibility of clouds, when comparing emission and transmission spectra, can be explained by the different geometries when probing the atmosphere in transmission \citep{fortney2005}. For transmission, we again find a good agreement between \petit and \ptrad.

For completeness, we also show a comparison to a model with even thicker clouds, see Figure \ref{fig:pt_rad_ver_very_cloudy} in Appendix \ref{sect:app_very_cloudy}, corresponding to Cloud Model 5 in \citet{mollierevanboekel2016}. In this model the cloud mass fraction is only capped at $10^{-2}\cdot Z_{\rm P}$. While we again find a good agreement between the transmission spectra of \petit and \ptrad, this is not the case for emission spectra, out to wavelengths of 4.5~$\mu$m, due to the strong scattering of light by cloud particles (resulting in both reflection of stellar light and scattering of light originating within the planet atmosphere).

\section{Retrieval examples, low resolution (\emph{`c-k'}) mode}
\label{sect:ref_test}
In this section we show an exemplary use of \ptrad \ for a retrieval analysis. Here, we focus on simple, clear atmosphere scenarios.

\begin{table}[t!]
\centering
{ \footnotesize
\begin{tabular}{ll}
\hline \hline
Parameter & Description \\ \hline
$X_i$ & Mass fraction of absorber species $i$ \\
$\mu$ & Mean molecular weight \\
$P_0$ & Reference pressure \\
$R_{\rm P}$ & Planet radius at $P_0$ \\
$g$ & Atmospheric surface gravity at $P_0$ \\
$\xi_j$ & Free parameters of the $T(P)$ profile \\
\hline
\end{tabular}
}
\caption{Parameters of the retrieval model used in this study. Not all parameters are necessarily free parameters, see Section \ref{sect:retrieval_setup}.}
\label{tab:ret_free_params}

\centering
{ \footnotesize
\begin{tabular}{ll|ll}
\hline \hline
Abundances & Value & Parameter & Value \\ \hline
$X_{\ce{CH4}}$ & $7.71\times 10^{-9}$ & $P_0$ & 0.01 bar \\
$X_{\ce{CO}}$ & $5.52\times 10^{-3}$ & $R_{\rm P}$ & 1.84 $\rj$\\
$X_{\ce{H2O}}$ & $2.46\times 10^{-3}$ & $g$ & 380 cm s$^{-2}$\\
$X_{\ce{H2S}}$ & $2.40\times 10^{-4}$ & $\alpha$ (i.e., $\xi_1$) & 0.5 \\
$X_{\ce{K}}$ & $1.52\times 10^{-6}$ & $P_{\rm trans}$ (i.e., $\xi_2$) & 0.001 bar\\
$X_{\ce{NH3}}$ & $3.80\times 10^{-8}$ & $\gamma$ (i.e., $\xi_3$)& 0.4 \\
$X_{\ce{Na}}$ & $2.46\times 10^{-5}$ & $\delta$ (i.e., $\xi_4$)& $10^{-5}$ bar$^{-1}$\\
$X_{\ce{CO2}}$ & $8.48\times 10^{-7}$ & $T_{\rm int}$ (i.e., $\xi_5$)& 600 K \\
 &  & $T_{\rm eq}$ (i.e., $\xi_6$)& 1900 K \\ \hline \hline
 Instrument & Noise floor & $T_{*}$ & 6295~K \\ \hline
 \emph{NIRISS SOSS} & 20 ppm & $R_*$ & 1.81 $\rsun$\\
 \emph{NIRSpec G395M} & 75 ppm & $K$ & 10.33 mag\\
 \emph{MIRI LRS} & 40 ppm & $t_{\rm trans}$ & 3.658 h\\
 & & $f_{\rm base}$ & 1 \\
\hline
\end{tabular}
}
\caption{Parameters for generating the clear TrES-4b-like model (top) and synthetic observations (bottom). $f_{\rm base}$ denotes the ratio of the out-of-transit and the in-transit observing time.}
\label{tab:tres_4_b}
\end{table}

\subsection{Retrieval setup}
\label{sect:retrieval_setup}
\ptrad \ is a radiative transfer code for Python. That means that the retrieval needs to be implemented by the user, with Python tools such as the MCMC implementation \emph{emcee} \citep{foreman-mackeyhogg2013}, or the nested sampling implementation \emph{PyMultiNest} \citep{buchnergeorgakakis2014}. In what follows below, we will construct a simple retrieval example using \emph{emcee}. This example setup is also published and documented in the manual of \ptrad. Users of \ptrad \ may use this as a starting point for their own retrievals, or construct independent setups themselves.

The model of the retrieval example shown here is defined by the parameters listed in Table \ref{tab:ret_free_params}. Not all of these are necessarily free parameters. For example, below the retrieval of a transiting planet is studied, for which the planetary mass $M_{\rm P}$, as well as a white-light radius $R_{\rm wl}$, is known. One can thus fix $g = GM_{\rm P}/R_{\rm P}^2$ and $R_{\rm P} = R_{\rm wl}$, leaving the reference pressure $P_0$ as a free parameter, to be found by the retrieval.

Here, vertically constant mass fractions are assumed. The abundances of \ce{H2} and He are assumed to be fixed through calculating
\beq
X_{\rm H \ and \ He} = 1-\sum_{i=1}^{N_{\rm metals}}X_i \ ,
\eeq
where $N_{\rm metals}$ is the number of metal species (all absorbers except \ce{H2} and He), and using this to calculate
\beq
X_{\rm H_2} = 0.75 \cdot X_{\rm H \ and \ He}
\eeq
and
\beq
X_{\rm He} = 0.25 \cdot X_{\rm H \ and \ He} .
\eeq
This assumes a gas of approximately primordial composition. Molecules such as H$_2$O are counted to belong to the metals, even though they also contain hydrogen.

\begin{figure*}[t!]
\centering
\includegraphics[width=0.8\textwidth]{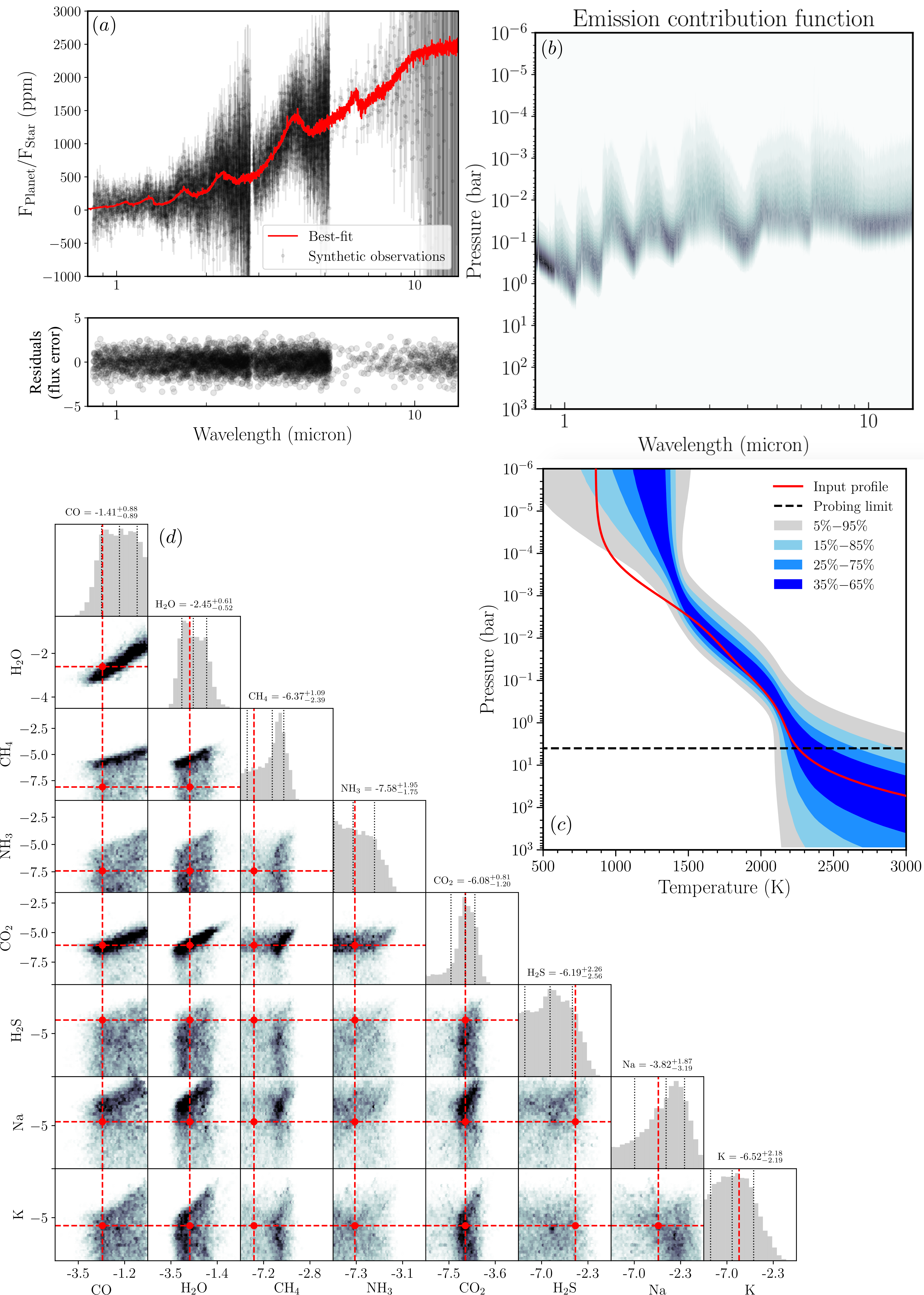}
\caption{Retrieval using mock \emph{JWST} emission spectra of a TrES-4b-like planet. Panel (a), top: synthetic observations (black circles) and best-fit spectrum (red solid line), Panel (a), bottom: residuals between the best-fit spectrum and mock observations, in units of the observational errors. Panel (b): emission contribution function, calculated using Equation \ref{equ:emis_contr}. Pressures higher than 4~bar cannot be probed. Panel (c): confidence envelopes of the retrieved $P$-$T$ profiles (light gray to dark blue) and input $P$-$T$ profile (red solid line). The last 600,000 samples were used for making this plot. The probing limit, inferred from Panel (b), is shown as a horizontal black dashed line. Below this altitude the $P$-$T$ profile cannot be probed, all posterior envelopes are determined by the analytic form of the $P$-$T$ function (Equation \ref{equ:retrieval_temp_model}). Panel (d): 2-d posterior plot of the retrieved log-mass fractions, input values are shown as red dashed lines. The last 700,000 samples were used for making this plot.}
\label{fig:emis_res}
\end{figure*}

The mean molecular weight $\mu$ is calculated through
\beq
\frac{1}{\mu} = \sum_{i=1}^{N_{\rm species}}\frac{X_i}{\mu_i},
\eeq
where $N_{\rm species} = N_{\rm metals}+2$, to account for \ce{H2} and He. In this example we hence assume that there are no non-absorbing chemical species that carry a significant amount of mass. \rch{We note that the mean molecular weight in general should not include condensed species, because they are not sufficiently coupled to the atmospheric gas, see \citet{baudinomolliere2017}.}

The temperature parametrization is given by
\beq
T(P) = \left<T_{\rm Guillot}(P)\cdot\left(1-\frac{\alpha}{1+P/P_{\rm trans}}\right)\right>_P \ ,
\label{equ:retrieval_temp_model}
\eeq
with
\begin{multline}
T_{\rm Guillot}(P) = \frac{3T_{\rm int}^4}{4}\left(\frac{2}{3}+\delta P\right) + \\
\frac{3T_{\rm eq}^4}{4}\left[\frac{2}{3}+\frac{1}{\gamma\sqrt{3}}+\left(\frac{\gamma}{\sqrt{3}}-\frac{1}{\gamma\sqrt{3}}\right)e^{-\gamma\delta\sqrt{3}P}\right] \ ,
\end{multline}
which is the \citet{guillot2010} temperature model. In the \citet{guillot2010} model the optical depth is defined as $\tau=P\kappa_{\rm IR}/g$. We defined the new parameter $\delta = \kappa_{\rm IR}/g$, such that $\tau=\delta \cdot P$. The $\left<\right>_P$ denotes a boxcar smoothing, carried out over a ${\rm log}(P)$-width of 1.25 dex. The temperature model thus has six free parameters, namely $\xi_1=\alpha$, $\xi_2=P_{\rm trans}$, $\xi_3=\gamma$, $\xi_4=\delta$, $\xi_5=T_{\rm int}$, and $\xi_6=T_{\rm eq}$.

The second term on the RHS of Equation \ref{equ:retrieval_temp_model} was introduced to allow the upper part of the atmosphere do develop a non-isothermal behavior at high altitudes, that is, low pressures. This alleviates that the \citet{guillot2010} temperature model without this additional term will always lead to isothermal atmospheres at high altitudes, because it assumes a (double-)gray opacity. This is not the case for non-gray atmospheric models, \citep[see, e.g.,][Figure 5]{mollierevanboekel2015}. The reason for this is discussed in Chapter 6.4.2 of \citet{molliere2017}.

We have found this temperature model to be sufficiently flexible when fitting it to a diverse set of self-consistent temperature profiles (clear, cloudy, with and without inversions), see Figure 6.2 in \citet{molliere2017}.

\subsection{Retrieval of synthetic spectra generated with \ptrad}

In this section the results of retrieving an atmospheric structure with \ptrad \ are shown. For the input spectrum to be retrieved, we created synthetic observations of a \ptrad \ spectrum, using parameters motivated by the exoplanet TrES-4b. TrES-4b should be an excellent target for \emph{JWST}, likely being in the temperature range to probe the cloud properties of hot condensates species such as silicates \citep{mollierevanboekel2016}. For the clear atmosphere retrieved here, the input parameters as reported in Table \ref{tab:tres_4_b} (upper part) were used. The resulting effective temperature of the synthetic emission spectrum is 1789~K, which is close to equilibrium temperature value of this planet, 1795~K \citep{sozzettibonomo2015}.

We produced single transit/eclipse synthetic observations of \emph{JWST}'s \emph{NIRISS SOSS}, \emph{NIRSpec G395M}, and \emph{MIRI LRS} modes using the \emph{PandExo} tool \citep[][]{batalhamandell2017}. For this we used the parameters given in the lower part of Table \ref{tab:tres_4_b}. The saturation level was assumed to be 80~\% of the full well capacity for all instrument modes. For the stellar spectrum necessary for calculating the planet-to-star contrast we used the library of \emph{PHOENIX} \citep{husserwende-vonberg2013} and \emph{ATLAS9} \citep{kurucz1979,kurucz1992,kurucz1994} spectra that is part of \ptrad. Here, the stellar flux is returned as a function of the stellar effective temperature, using the model grid from \citet{vanboekel2012}.

\begin{table}[t!]
\centering
{ \footnotesize
\begin{tabular}{llll}
\hline \hline
Parameter & Setup & Prior & Additional prior \\ \hline
${\rm log}(X_i)$ & $\mathcal{U}(-10,0)$ & $\mathcal{U}(-10,0)$ & $\sum_iX_i\leq1$ \\
${\rm log}(\delta)$ & $\mathcal{N}(-5.5,2.5)$ & $\mathcal{N}(-5.5,2.5)$ & \\
${\rm log}(\gamma)$ & $\mathcal{N}(0,2)$ & $\mathcal{N}(0,2)$ &  \\
$T_{\rm int}$ & $\mathcal{U}(0,1500)$ & $\mathcal{U}(0,1500)$ &  \\
$T_{\rm eq}$ & $\mathcal{U}(0,4000)$ & $\mathcal{U}(0,4000)$ &  \\
${\rm log}(P_{\rm trans})$ & $\mathcal{N}(-3,3)$ & $\mathcal{N}(-3,3)$ &  \\
$\alpha$ & $\mathcal{N}(0.25,0.4)$ & $\mathcal{N}(0.25,0.4)$ & $\alpha<1$  \\
\hline
\end{tabular}
}
\caption{Distributions used to sample the walker positions and prior choice. $\mathcal{U}$ stands for a uniform distribution, with the two parameters being the range boundaries. $\mathcal{N}$ stands for a normal distribution, with the two parameters being the mean value and standard deviation. The unit of $\delta$ is bar$^{-1}$, the unit of $T_{\rm int}$ and $T_{\rm eq}$ is K, and the unit of $P_{\rm trans}$ is bar.}
\label{tab:samp_prior}
\end{table}

\subsubsection{Emission spectra}
We set the \emph{emcee} MCMC sampler up with 240 walkers. They were initialized using the distributions described in Table \ref{tab:samp_prior}. For most of the parameters it made sense to retrieve their logarithm, this is indicated in Table \ref{tab:samp_prior} as well. Table \ref{tab:samp_prior} also describes which distributions were used for the priors during the retrieval; they are identical to the distributions sampled for the walker setup. The parameter values of the temperature profile priors were guided by fitting our temperature model (Equation \ref{equ:retrieval_temp_model}) to the temperature structures published in \citet{mollierevanboekel2016}. \citet{mollierevanboekel2016} contains a selection of self-consistent models (clear and cloudy) for transiting planets.

\rcht{We note that we allow maximum values of 4000~K for the $T_{\rm eq}$ parameter of the temperature model (Equation \ref{equ:retrieval_temp_model}), even though the line opacity database only goes up to 3000~K, after which the respective opacities are taken to be constant at their 3000~K value. However, in the retrieval example of the hot Jupiter presented in this study (TrES-4b, with $T_{\rm eq}\sim 1800$~K), the deepest atmospheric layers, which are hotter than 3000~K, are not probed by outgoing radiation or the grazing stellar rays (see contribution plots and probing limits in Panels b and c in Figures \ref{fig:emis_res} and \ref{fig:transm_res} below). In addition, keeping the parameters of the temperature model flexible (e.g., $T_{\rm eq}$ up to 4000~K) is useful for letting the retrieval find the $P$-$T$ parameters that best describe the data. If the $\alpha$ parameter (see Equation \ref{equ:retrieval_temp_model}) is very large (but less than one), for example, it can strongly decrease the atmospheric temperature, even at high $T_{\rm eq}$ values. Lastly, while the line opacities are constant for $T>3000$~K, the radiative source function, and the atmospheric scale height, are not. So if the $T_{\rm eq}>3000$~K choice may lead to situations where regions hotter than 3000~K are sampled by the outgoing flux, this flux will still be high, and the transit radius large. Thus, the retrieval may ``tell'' the user that it retrieves temperatures in excess of 3000~K, from which point on one would have to be extremely careful in one's interpretation of the retrieval results.}

100,000 samples were drawn for the pre-burn run, then a chain was started to draw 1,000,000 samples, with the walker positions initialized in a Gauss ball around the best-fit positions of the pre-burn. As a second test we reran the retrieval starting around the median position of the first chain, and drew 1,000,000 samples, with the same outcome.

Figure \ref{fig:emis_res} summarizes the results from retrieving the properties of the TrES-4b-like planet with a mock \emph{JWST} emission spectrum. Panel (a) shows the input synthetic observations and the best-fit spectrum. The fit is excellent, this is further corroborated by the residual plot of Panel (a), which shows that the residuals scatter about zero, with most values within 2~$\sigma$.

Panel (b) of Figure \ref{fig:emis_res} shows the emission contribution function of the best-fit spectrum, which was calculated using Equation \ref{equ:emis_contr}. One can see that pressures larger than 4~bar cannot be probed. The temperature distribution at lower altitudes arises from the analytic shape of the temperature profile.

Panel (c) shows the distribution of retrieved temperature profiles and the input profile. It also displays the probing limit (4~bar), as derived from Panel (b). Between $10^{-3}$ and 4 bar, the $P$-$T$ envelopes closely follow the position and slope of the input profile. This is also where most of the spectrum contributes, see Panel (b). At lower pressure the retrieval fails to derive the temperatures correctly; here the emission contribution function is very small. At larger pressures the temperature appears to be well constrained. Because this high pressure region cannot be probed, its retrieved temperature is fully determined by the predictions of the temperature model.

Panel (d) displays the 2-d and 1-d projections of the posterior distribution. The abundances of the major absorbers CO and \ce{H2O} are retrieved within the 16-84~\% percentile boundaries of the 1-d posteriors. If the 1-d posteriors were identical to a Gaussian distribution these percentile boundaries would correspond to the 1-$\sigma$ error bars.

For all the other species, that is for \ce{CH4}, \ce{NH3}, \ce{CO2}, \ce{H2S}, Na and K, only upper limits can be retrieved. The peak value of the \ce{CO2} mass fraction is almost identical to the input value, but there exists a signifiant probability tail toward lower abundances. For Na and K quite large upper limits are retrieved for the mass fractions, which is possible especially as they can only affect the spectrum at the blue end of the spectrum at 0.8 micron, where only little flux escapes the planet. Moreover, the strong doublet lines of Na and K do not fall within the wavelength range of the synthetic JWST observations, and hence they can only affect the spectrum by virtue of their line wings. 

\begin{figure*}[t!]
\centering
\includegraphics[width=0.8\textwidth]{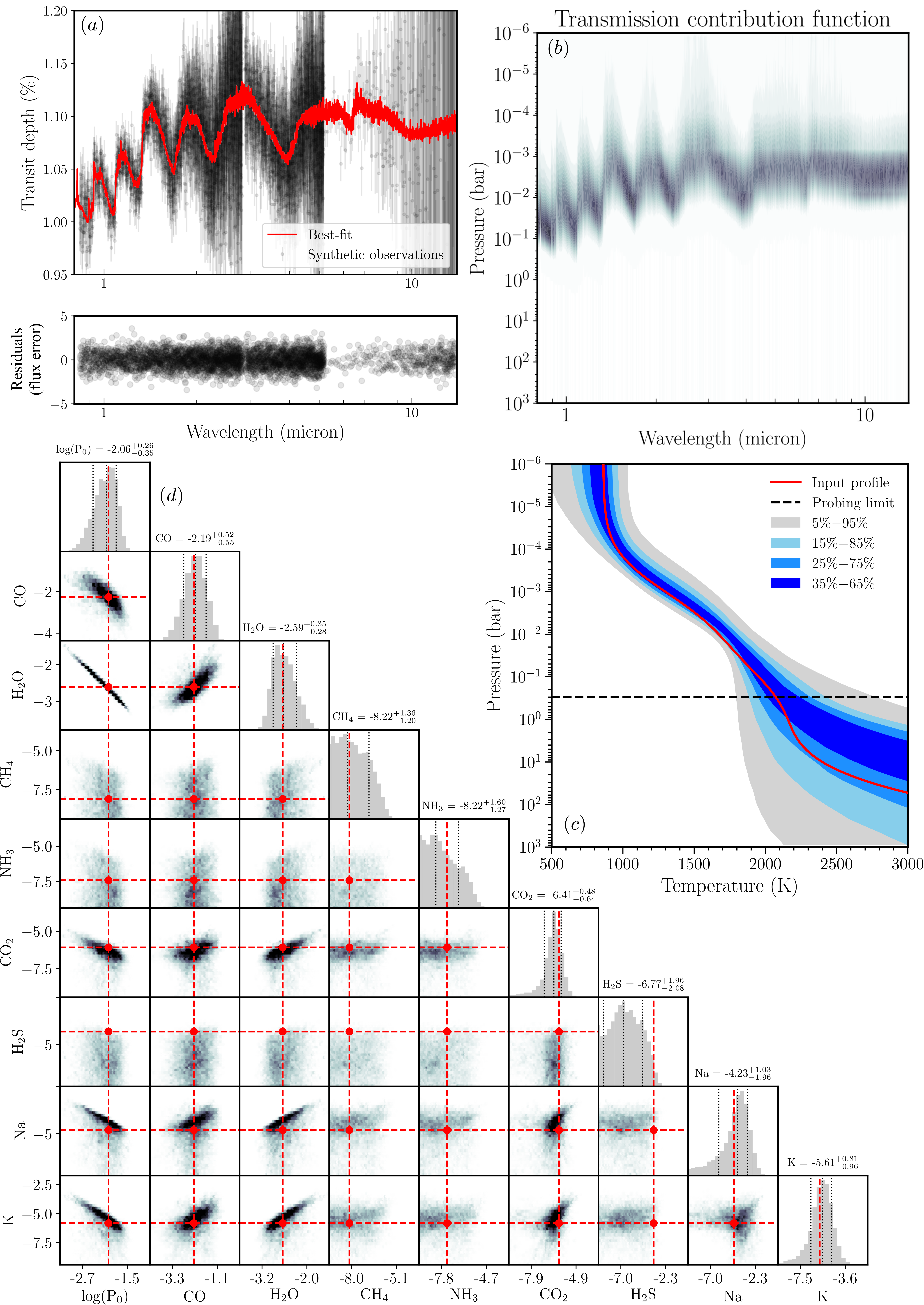}
\caption{Same as Figure \ref{fig:emis_res}, but for the retrieval of the TrES-4b {\it transmission} spectrum.}
\label{fig:transm_res}
\end{figure*}

\subsubsection{Transmission spectra}

In this section we describe the results from retrieving the properties of the TrES-4b-like planet with a mock \emph{JWST} transmission spectrum.
We again used 240 walkers, a pre-burn run with 100,000 samples, and then drew 1,000,000 samples, initializing the walkers around the best-fit position of the pre-burn. Figure \ref{fig:transm_res} summarizes the results of the retrieval.

In Panel (a) the input synthetic observations are shown and the best-fit spectrum of the retrieval. The fit appears to be excellent, just as in the case of the emission spectrum retrieval. The residuals scatter about zero again (see lower panel of Panel a), with most values within 2~$\sigma$.

Panel (b) of Figure \ref{fig:transm_res} shows the transmission contribution function for the best-fit spectrum, calculated using Equation \ref{equ:trans_contr}. One sees that pressures larger than 0.3~bar cannot be probed, and that all temperature information at lower altitudes arises from the analytic shape of the temperature profile, and the parameter distribution of the temperature profile, which is sensitive to pressures smaller than 0.3~bar.

Panel (c) shows the distribution of retrieved temperature profiles, the input profile, as well as the probing limit (0.3~bar), which is derived from Panel (b). The $P$-$T$ envelopes closely follow the position and slope of the  input profile. The largest deviation is at $P\sim 0.01$~bar, where the input temperature profile only falls within the 15 to 85~\% envelope (i.e., still within the ``1-$\sigma$ range'').

Finally, the 2 and 1-d projection of the full posterior is shown in Panel (d) of Figure \ref{fig:transm_res}, for all other parameters of interest; namely the reference pressure $P_0$, which is retrieved very well, and the log-mass fractions of all atmospheric absorbers.

For the absorbers with high abundance in the input model, that is, CO and \ce{H2O}, the abundances are correctly retrieved, that is the input value falls within the 15 to 85~\% percentile values (``1-$\sigma$ range'') of the retrieved values. This also holds for \ce{CO2}, Na, and K, although they are of much lower abundance. Their 1-d posteriors also show a tail toward lower abundances. Yet, the good constraints on especially Na and K are surprising, because the spectral range studied here is only sensitive to the quasi-continuum absorption of the strong line wings of optical doublets of Na and K, the exact shape and strength of which are uncertain \citep[see the discussion in, e.g.,][]{sharpburrows2007,baudinomolliere2017}. The wing absorption of Na and K at the short \emph{JWST} wavelengths are also likely degenerate with condensate absorption in cloudy atmospheres in retrievals of real observations. Hence the power of the \emph{JWST} wavelength range for inferring Na and K abundances is probably low. For the three species with lowest abundance (\ce{CH4}, \ce{NH3}, and \ce{H2S}) only upper limits are retrieved.

\section{Summary}
\label{sect:summary}

In this paper, we report on an easy-to-use Python package for the synthesis of exoplanet spectra, called \ptrad.
\ptrad \ allows the user to calculate spectra using a multitude of different modes. Two different resolutions are available. The low-resolution mode, at $\lambda / \Delta \lambda = 1000$, makes use of the correlated-k assumption, and excellently agrees with the results of the high-resolution mode ($\lambda / \Delta \lambda = 10^6$), which produces spectra with line-by-line radiative transfer. The differences in flux are usually below 1~\%, translating to errors of less than 10~ppm for planet-to-star flux contrasts of $10^{-3}$, and less for fainter planets. Both emission and transmission spectra can be calculated. Moreover, the code allows to include clouds, using either parametric descriptions for the wavelength dependence of the cloud opacity, or cross-sections derived from optical constants of various condensates. For the latter spherical or irregular particle shapes can be assumed, and often cross-sections both for crystalline and amorphous particles are available. The high resolution opacities are often provided for individual isotopologue species, see the code documentation website\footnote{\ptdoc} for the up-to-date list of opacities. \rch{The documentation website also contains a short tutorial of the \emph{ExoCross} code \citep{yurchenkoalrefaie2018}, and how to convert its resulting \emph{Exomol} opacity calculations for use in \ptrad,} \rcht{allowing to calculate line opacities of additional species.}

We successfully verify \ptrad \ by comparing to \emph{petitCODE}, which is itself a well-tested and benchmarked code \citep{baudinomolliere2017}, having been used in the literature for exoplanet characterization, for example via grid retrieval \citep{samlandmolliere2017}. \rch{Differences between \petit and \ptrad \ are usually below 1~\%, but can be of order 5~\% in regions of appreiable flux. The latter translates to errors of up to 50 ppm for planet-to-star flux contrasts of $10^{-3}$, and less for fainter planets. The error bars of \emph{JWST} observations, arising from pure photon noise, can be on the order of 10 ppm, when stacking multiple observations. However, systematic noise is expected to affect \emph{JWST} observations, with noise value estimates ranging from 20 to 50 ppm \citep[see, e.g.,][]{greeneline2016}. The differences between \emph{petitCODE} and \emph{petitRADTRANS} are comparable with these noise floor values.}

We show two retrieval examples, applying \ptrad \ to synthetic \emph{JWST} emission and transmission spectra. The retrievals were carried out using cloud-free input models, drawing $10^6$ samples each, successfully constraining the planetary temperature structure and abundances.

\ptrad \ spectra are calculated within a few seconds at both low and high resolution. For the high resolution mode, this describes the computational time over wavelength ranges typically covered by high resolution observations. We decided against sacrificing accuracy for computational speed, and employ the correlated-k method at low resolution, rather than opacity sampling. Opacity sampling would make \ptrad \ at least 16 times faster, but leads to white noise at the 100 ppm level for the transit depths, see \citet{zhangchachan2019}. The retrievals presented here (calculating 10$^6$ spectra each) were carried out on 30 cores on the \emph{Batchelor} cluster at the Max Planck Institute for Astronomy in Heidelberg, Germany, and finished within a few days.

\ptrad \ is a radiative transfer package well suited for carrying out retrievals. The commented retrieval setups \citep[using \emph{emcee}, see][]{foreman-mackeyhogg2013} are available from the \ptrad \ documentation website. Although the retrieval cases presented here are simple, adding complexity to the retrieval model is easily possible, for example by mixing clear and cloudy radiative transfer objects for patchy cloud cases, or using the \emph{PHOENIX} spectral library that is part of \ptrad \ to account for surface inhomogeneities of the host star, using Equation 1 of \citet{rackhamapai2018}. Moreover, with its high-resolution mode, \ptrad \ also allows to carry out high-resolution retrievals, as recently advertized by \citet{brogiline2018}.

The versatility of the \ptrad \ package makes it an ideal and timely tool for exoplanet characterization and retrieval studies. We plan to update its capabilities and opacity database frequently, enabling users to do exciting science for the years to come.

\begin{acknowledgements}
We thank our anonymous referee for his or her comments, which improved the quality of this paper. P.M. thanks M. Rocchetto for insightful discussions about the correlated-k assumption. He also thanks J. Wang for helpful comments on this manuscript. P.M. thanks H. Klahr for his continued computational hospitality on the \emph{Batchelor} cluster. P.M. is thankful to K. Chubb for an ExoCross tutorial. P.M and I.S. acknowledge support from the European Research Council under the European Union's Horizon 2020 research and innovation program under grant agreement No. 694513.
\end{acknowledgements}

\bibliographystyle{aa}
\bibliography{mybib}{}

\newcommand{\noop}[1]{}
\begin{thebibliography}{111}
\expandafter\ifx\csname natexlab\endcsname\relax\def\natexlab#1{#1}\fi

\bibitem[{{Ackerman} \& {Marley}(2001)}]{ackermanmarley2001}
{Ackerman}, A.~S. \& {Marley}, M.~S. 2001, \apj, 556, 872

\bibitem[{{Arcangeli} {et~al.}(2018){Arcangeli}, {D{\'e}sert}, {Line}, {Bean},
  {Parmentier}, {Stevenson}, {Kreidberg}, {Fortney}, {Mansfield}, \&
  {Showman}}]{arcangelidesert2018}
{Arcangeli}, J., {D{\'e}sert}, J.-M., {Line}, M.~R., {et~al.} 2018, \apj, 855,
  L30

\bibitem[{{Barber} {et~al.}(2014){Barber}, {Strange}, {Hill}, {Polyansky},
  {Mellau}, {Yurchenko}, \& {Tennyson}}]{barber2014}
{Barber}, R.~J., {Strange}, J.~K., {Hill}, C., {et~al.} 2014, \mnras, 437, 1828

\bibitem[{{Barton} {et~al.}(2013){Barton}, {Yurchenko}, \&
  {Tennyson}}]{bartonyurchenko2013}
{Barton}, E.~J., {Yurchenko}, S.~N., \& {Tennyson}, J. 2013, \mnras, 434, 1469

\bibitem[{{Batalha} {et~al.}(2017){Batalha}, {Mandell}, {Pontoppidan},
  {Stevenson}, {Lewis}, {Kalirai}, {Earl}, {Greene}, {Albert}, \&
  {Nielsen}}]{batalhamandell2017}
{Batalha}, N.~E., {Mandell}, A., {Pontoppidan}, K., {et~al.} 2017, \pasp, 129,
  064501

\bibitem[{{Baudino} {et~al.}(2015){Baudino}, {B{\'e}zard}, {Boccaletti},
  {Bonnefoy}, {Lagrange}, \& {Galicher}}]{baudinobezard2015}
{Baudino}, J.-L., {B{\'e}zard}, B., {Boccaletti}, A., {et~al.} 2015, \aap, 582,
  A83

\bibitem[{{Baudino} {et~al.}(2017){Baudino}, {Molli{\`e}re}, {Venot},
  {Tremblin}, {B{\'e}zard}, \& {Lagage}}]{baudinomolliere2017}
{Baudino}, J.-L., {Molli{\`e}re}, P., {Venot}, O., {et~al.} 2017, \apj, 850,
  150

\bibitem[{{Beichman} {et~al.}(2014){Beichman}, {Benneke}, {Knutson}, {Smith},
  {Lagage}, {Dressing}, {Latham}, {Lunine}, {Birkmann}, {Ferruit}, {Giardino},
  {Kempton}, {Carey}, {Krick}, {Deroo}, {Mandell}, {Ressler}, {Shporer},
  {Swain}, {Vasisht}, {Ricker}, {Bouwman}, {Crossfield}, {Greene}, {Howell},
  {Christiansen}, {Ciardi}, {Clampin}, {Greenhouse}, {Sozzetti}, {Goudfrooij},
  {Hines}, {Keyes}, {Lee}, {McCullough}, {Robberto}, {Stansberry}, {Valenti},
  {Rieke}, {Rieke}, {Fortney}, {Bean}, {Kreidberg}, {Ehrenreich}, {Deming},
  {Albert}, {Doyon}, \& {Sing}}]{beichmanbenneke2014}
{Beichman}, C., {Benneke}, B., {Knutson}, H., {et~al.} 2014, \pasp, 126, 1134

\bibitem[{{Benneke}(2015)}]{benneke2015}
{Benneke}, B. 2015, ArXiv e-prints

\bibitem[{{Benneke} \& {Seager}(2012)}]{bennekeseager2012}
{Benneke}, B. \& {Seager}, S. 2012, \apj, 753, 100

\bibitem[{{Blecic} {et~al.}(2017){Blecic}, {Dobbs-Dixon}, \&
  {Greene}}]{blecicdobbsdixon2017}
{Blecic}, J., {Dobbs-Dixon}, I., \& {Greene}, T. 2017, \apj, 848, 127

\bibitem[{{Borysow}(2002)}]{borysow2002}
{Borysow}, A. 2002, \aap, 390, 779

\bibitem[{{Borysow} \& {Frommhold}(1989)}]{borysow1989b}
{Borysow}, A. \& {Frommhold}, L. 1989, \apj, 341, 549

\bibitem[{{Borysow} {et~al.}(1989){Borysow}, {Frommhold}, \&
  {Moraldi}}]{borysow1989a}
{Borysow}, A., {Frommhold}, L., \& {Moraldi}, M. 1989, \apj, 336, 495

\bibitem[{{Borysow} {et~al.}(2001){Borysow}, {Jorgensen}, \&
  {Fu}}]{borysow2001}
{Borysow}, A., {Jorgensen}, U.~G., \& {Fu}, Y. 2001, \jqsrt, 68, 235

\bibitem[{{Borysow} {et~al.}(1988){Borysow}, {Frommhold}, \&
  {Birnbaum}}]{borysow1988}
{Borysow}, J., {Frommhold}, L., \& {Birnbaum}, G. 1988, \apj, 326, 509

\bibitem[{{Brogi} {et~al.}(2016){Brogi}, {Line}, {Bean}, {D{\'e}sert}, \&
  {Schwarz}}]{brogiline2016}
{Brogi}, M., {Line}, M., {Bean}, J., {D{\'e}sert}, J.-M., \& {Schwarz}, H.
  2016, ArXiv e-prints

\bibitem[{{Brogi} \& {Line}(2018)}]{brogiline2018}
{Brogi}, M. \& {Line}, M.~R. 2018, arXiv e-prints

\bibitem[{{Bryan} {et~al.}(2018){Bryan}, {Benneke}, {Knutson}, {Batygin}, \&
  {Bowler}}]{bryanbenneke2018}
{Bryan}, M.~L., {Benneke}, B., {Knutson}, H.~A., {Batygin}, K., \& {Bowler},
  B.~P. 2018, Nature Astronomy, 2, 138

\bibitem[{{Buchner} {et~al.}(2014){Buchner}, {Georgakakis}, {Nandra}, {Hsu},
  {Rangel}, {Brightman}, {Merloni}, {Salvato}, {Donley}, \&
  {Kocevski}}]{buchnergeorgakakis2014}
{Buchner}, J., {Georgakakis}, A., {Nandra}, K., {et~al.} 2014, \aap, 564, A125

\bibitem[{Burch {et~al.}(1969)Burch, Gryvnak, Patty, \&
  Bartky}]{burchgryvnak1969}
Burch, D.~E., Gryvnak, D.~A., Patty, R.~R., \& Bartky, C.~E. 1969, J. Opt. Soc.
  Am., 59, 267

\bibitem[{Chan \& Dalgarno(1965)}]{chandalgarno1965}
Chan, Y.~M. \& Dalgarno, A. 1965, Proceedings of the Physical Society, 85, 227

\bibitem[{{Crossfield} {et~al.}(2014){Crossfield}, {Biller}, {Schlieder},
  {Deacon}, {Bonnefoy}, {Homeier}, {Allard}, {Buenzli}, {Henning}, {Brandner},
  {Goldman}, \& {Kopytova}}]{crossfieldbiller2014}
{Crossfield}, I.~J.~M., {Biller}, B., {Schlieder}, J.~E., {et~al.} 2014, \nat,
  505, 654

\bibitem[{{Cubillos}(2016)}]{cubillos2017}
{Cubillos}, P.~E. 2016, arXiv e-prints, arXiv:1604.01320

\bibitem[{{Dalgarno} \& {Williams}(1962)}]{dalgarnowilliams1962}
{Dalgarno}, A. \& {Williams}, D.~A. 1962, \apj, 136, 690

\bibitem[{{Feng} {et~al.}(2016){Feng}, {Line}, {Fortney}, {Stevenson}, {Bean},
  {Kreidberg}, \& {Parmentier}}]{fengline2016}
{Feng}, Y.~K., {Line}, M.~R., {Fortney}, J.~J., {et~al.} 2016, \apj, 829, 52

\bibitem[{{Fischer} {et~al.}(2003){Fischer}, {Gamache}, {Goldman}, {Rothman},
  \& {Perrin}}]{fischer2003}
{Fischer}, J., {Gamache}, R.~R., {Goldman}, A., {Rothman}, L.~S., \& {Perrin},
  A. 2003, \jqsrt, 82, 401

\bibitem[{{Fisher} \& {Heng}(2018)}]{fisherheng2018}
{Fisher}, C. \& {Heng}, K. 2018, \mnras, 481, 4698

\bibitem[{{Flowers} {et~al.}(2018){Flowers}, {Brogi}, {Rauscher}, {M-R
  Kempton}, \& {Chiavassa}}]{flowersbrogi2018}
{Flowers}, E., {Brogi}, M., {Rauscher}, E., {M-R Kempton}, E., \& {Chiavassa},
  A. 2018, arXiv e-prints

\bibitem[{{Foreman-Mackey} {et~al.}(2013){Foreman-Mackey}, {Hogg}, {Lang}, \&
  {Goodman}}]{foreman-mackeyhogg2013}
{Foreman-Mackey}, D., {Hogg}, D.~W., {Lang}, D., \& {Goodman}, J. 2013, \pasp,
  125, 306

\bibitem[{{Fortney}(2005)}]{fortney2005}
{Fortney}, J.~J. 2005, \mnras, 364, 649

\bibitem[{{Fu} \& {Liou}(1992)}]{fuliou1998}
{Fu}, Q. \& {Liou}, K.~N. 1992, Journal of Atmospheric Sciences, 49, 2139

\bibitem[{{Garland} \& {Irwin}(2019)}]{garlandirwin2019}
{Garland}, R. \& {Irwin}, P.~G.~J. 2019, arXiv e-prints, arXiv:1903.03997

\bibitem[{{Gharib-Nezhad} \& {Line}(2018)}]{gharibline2018}
{Gharib-Nezhad}, E. \& {Line}, M.~R. 2018, ArXiv e-prints

\bibitem[{{Gray}(2008)}]{gray08}
{Gray}, D.~F. 2008, {The Observation and Analysis of Stellar Photospheres}

\bibitem[{{Greene} {et~al.}(2016){Greene}, {Line}, {Montero}, {Fortney},
  {Lustig-Yaeger}, \& {Luther}}]{greeneline2016}
{Greene}, T.~P., {Line}, M.~R., {Montero}, C., {et~al.} 2016, \apj, 817, 17

\bibitem[{{Guillot}(2010)}]{guillot2010}
{Guillot}, T. 2010, \aap, 520, A27

\bibitem[{{Harris} {et~al.}(2006){Harris}, {Tennyson}, {Kaminsky}, {Pavlenko},
  \& {Jones}}]{harris2006}
{Harris}, G.~J., {Tennyson}, J., {Kaminsky}, B.~M., {Pavlenko}, Y.~V., \&
  {Jones}, H.~R.~A. 2006, \mnras, 367, 400

\bibitem[{{Hartmann} {et~al.}(2002){Hartmann}, {Boulet}, {Brodbeck}, {van
  Thanh}, {Fouchet}, \& {Drossart}}]{hartmannboulet2002}
{Hartmann}, J.-M., {Boulet}, C., {Brodbeck}, C., {et~al.} 2002, J. Quant. Spec.
  Radiat. Transf., 72, 117

\bibitem[{{Harvey} {et~al.}(1998){Harvey}, {Gallagher}, \& {Levelt
  Sengers}}]{harveygallagher1998}
{Harvey}, A.~H., {Gallagher}, J.~S., \& {Levelt Sengers}, J.~M.~H. 1998,
  Journal of Physical and Chemical Reference Data, 27, 761

\bibitem[{{Henning} \& {Stognienko}(1996)}]{henningstognienko1996}
{Henning}, T. \& {Stognienko}, R. 1996, \aap, 311, 291

\bibitem[{{Husser} {et~al.}(2013){Husser}, {Wende-von Berg}, {Dreizler},
  {Homeier}, {Reiners}, {Barman}, \& {Hauschildt}}]{husserwende-vonberg2013}
{Husser}, T.-O., {Wende-von Berg}, S., {Dreizler}, S., {et~al.} 2013, \aap,
  553, A6

\bibitem[{{Irwin} {et~al.}(2008){Irwin}, {Teanby}, {de Kok}, {Fletcher},
  {Howett}, {Tsang}, {Wilson}, {Calcutt}, {Nixon}, \&
  {Parrish}}]{irwinteanby2008}
{Irwin}, P.~G.~J., {Teanby}, N.~A., {de Kok}, R., {et~al.} 2008, \jqsrt, 109,
  1136

\bibitem[{{Jaeger} {et~al.}(1998){Jaeger}, {Molster}, {Dorschner}, {Henning},
  {Mutschke}, \& {Waters}}]{jaegermolster1998}
{Jaeger}, C., {Molster}, F.~J., {Dorschner}, J., {et~al.} 1998, \aap, 339, 904

\bibitem[{{Koike} {et~al.}(1995){Koike}, {Kaito}, {Yamamoto}, {Shibai},
  {Kimura}, \& {Suto}}]{koikekaito1995}
{Koike}, C., {Kaito}, C., {Yamamoto}, T., {et~al.} 1995, \icarus, 114, 203

\bibitem[{{Kurucz}(1993)}]{kurucz1993}
{Kurucz}, R. 1993, SYNTHE Spectrum Synthesis Programs and Line Data.~Kurucz
  CD-ROM No.~18.~Cambridge, Mass.: Smithsonian Astrophysical Observatory,
  1993., 18

\bibitem[{{Kurucz}(1994)}]{kurucz1994}
{Kurucz}, R. 1994, Solar abundance model atmospheres for 0,1,2,4,8 km/s.~Kurucz
  CD-ROM No.~19.~ Cambridge, Mass.: Smithsonian Astrophysical Observatory,
  1994., 19

\bibitem[{{Kurucz}(1979)}]{kurucz1979}
{Kurucz}, R.~L. 1979, \apjs, 40, 1

\bibitem[{{Kurucz}(1992)}]{kurucz1992}
{Kurucz}, R.~L. 1992, in IAU Symposium, Vol. 149, The Stellar Populations of
  Galaxies, ed. B.~{Barbuy} \& A.~{Renzini}, 225

\bibitem[{{Lacis} \& {Oinas}(1991)}]{lacis_oinas1991}
{Lacis}, A.~A. \& {Oinas}, V. 1991, \jgr, 96, 9027

\bibitem[{{Lavie} {et~al.}(2017){Lavie}, {Mendon{\c c}a}, {Mordasini}, {Malik},
  {Bonnefoy}, {Demory}, {Oreshenko}, {Grimm}, {Ehrenreich}, \&
  {Heng}}]{laviemendoca2017}
{Lavie}, B., {Mendon{\c c}a}, J.~M., {Mordasini}, C., {et~al.} 2017, \aj, 154,
  91

\bibitem[{{Lee} {et~al.}(2012){Lee}, {Fletcher}, \& {Irwin}}]{leefletcher2012}
{Lee}, J.-M., {Fletcher}, L.~N., \& {Irwin}, P.~G.~J. 2012, \mnras, 420, 170

\bibitem[{{Lee} {et~al.}(2014){Lee}, {Irwin}, {Fletcher}, {Heng}, \&
  {Barstow}}]{leeirwin2014}
{Lee}, J.-M., {Irwin}, P.~G.~J., {Fletcher}, L.~N., {Heng}, K., \& {Barstow},
  J.~K. 2014, \apj, 789, 14

\bibitem[{{Line} {et~al.}(2014){Line}, {Knutson}, {Wolf}, \&
  {Yung}}]{lineknutson2014}
{Line}, M.~R., {Knutson}, H., {Wolf}, A.~S., \& {Yung}, Y.~L. 2014, \apj, 783,
  70

\bibitem[{{Line} {et~al.}(2016){Line}, {Marley}, {Liu}, {Morley}, {Burningham},
  {Hinkel}, {Teske}, \& {Fortney}}]{linemarley2016}
{Line}, M.~R., {Marley}, M.~S., {Liu}, M.~C., {et~al.} 2016, ArXiv e-prints

\bibitem[{{Line} \& {Parmentier}(2016)}]{lineparmentier2016}
{Line}, M.~R. \& {Parmentier}, V. 2016, \apj, 820, 78

\bibitem[{{Line} {et~al.}(2015){Line}, {Teske}, {Burningham}, {Fortney}, \&
  {Marley}}]{lineteske2015}
{Line}, M.~R., {Teske}, J., {Burningham}, B., {Fortney}, J.~J., \& {Marley},
  M.~S. 2015, \apj, 807, 183

\bibitem[{{Line} {et~al.}(2013){Line}, {Wolf}, {Zhang}, {Knutson}, {Kammer},
  {Ellison}, {Deroo}, {Crisp}, \& {Yung}}]{linewolf2013}
{Line}, M.~R., {Wolf}, A.~S., {Zhang}, X., {et~al.} 2013, \apj, 775, 137

\bibitem[{{Line} \& {Yung}(2013)}]{lineyung2013}
{Line}, M.~R. \& {Yung}, Y.~L. 2013, \apj, 779, 3

\bibitem[{{Line} {et~al.}(2012){Line}, {Zhang}, {Vasisht}, {Natraj}, {Chen}, \&
  {Yung}}]{linezhang2012}
{Line}, M.~R., {Zhang}, X., {Vasisht}, G., {et~al.} 2012, \apj, 749, 93

\bibitem[{{Lothringer} {et~al.}(2018){Lothringer}, {Barman}, \&
  {Koskinen}}]{lothringerbarman2018}
{Lothringer}, J.~D., {Barman}, T., \& {Koskinen}, T. 2018, \apj, 866, 27

\bibitem[{{Lothringer} \& {Barman}(2019)}]{lothringerbarman2019}
{Lothringer}, J.~D. \& {Barman}, T.~S. 2019, arXiv e-prints

\bibitem[{{MacDonald} \& {Madhusudhan}(2017)}]{macdonaldmadhusudhan2017}
{MacDonald}, R.~J. \& {Madhusudhan}, N. 2017, ArXiv e-prints

\bibitem[{{Madhusudhan} {et~al.}(2011){Madhusudhan}, {Harrington}, {Stevenson},
  {Nymeyer}, {Campo}, {Wheatley}, {Deming}, {Blecic}, {Hardy}, {Lust},
  {Anderson}, {Collier-Cameron}, {Britt}, {Bowman}, {Hebb}, {Hellier},
  {Maxted}, {Pollacco}, \& {West}}]{madhusudhan2011}
{Madhusudhan}, N., {Harrington}, J., {Stevenson}, K.~B., {et~al.} 2011, \nat,
  469, 64

\bibitem[{{Madhusudhan} \& {Seager}(2009)}]{madhusudhanseager2009}
{Madhusudhan}, N. \& {Seager}, S. 2009, \apj, 707, 24

\bibitem[{{Min} {et~al.}(2005){Min}, {Hovenier}, \& {de
  Koter}}]{minhovenier2005}
{Min}, M., {Hovenier}, J.~W., \& {de Koter}, A. 2005, \aap, 432, 909

\bibitem[{{Molli{\`e}re} \& {Snellen}(2018)}]{mollieresnellen2018}
{Molli{\`e}re}, P. \& {Snellen}, I.~A.~G. 2018, ArXiv e-prints

\bibitem[{{Molli{\`e}re} {et~al.}(2017){Molli{\`e}re}, {van Boekel}, {Bouwman},
  {Henning}, {Lagage}, \& {Min}}]{mollierevanboekel2016}
{Molli{\`e}re}, P., {van Boekel}, R., {Bouwman}, J., {et~al.} 2017, \aap, 600,
  A10

\bibitem[{{Molli{\`e}re} {et~al.}(2015){Molli{\`e}re}, {van Boekel},
  {Dullemond}, {Henning}, \& {Mordasini}}]{mollierevanboekel2015}
{Molli{\`e}re}, P., {van Boekel}, R., {Dullemond}, C., {Henning}, T., \&
  {Mordasini}, C. 2015, \apj, 813, 47

\bibitem[{Molli\`ere(2017)}]{molliere2017}
Molli\`ere, P.~M. 2017

\bibitem[{{Morley} {et~al.}(2012){Morley}, {Fortney}, {Marley}, {Visscher},
  {Saumon}, \& {Leggett}}]{morleyfortney2012}
{Morley}, C.~V., {Fortney}, J.~J., {Marley}, M.~S., {et~al.} 2012, \apj, 756,
  172

\bibitem[{Oliphant(2006)}]{numpy}
Oliphant, T. 2006, {NumPy}: A guide to {NumPy}, USA: Trelgol Publishing,
  [Online; accessed <today>]

\bibitem[{Palik(2012)}]{palik2012}
Palik, E. 2012, Handbook of Optical Constants of Solids No. Bd. 1 (Elsevier
  Science)

\bibitem[{{P\'egouri\'e}(1988)}]{pegourie1988}
{P\'egouri\'e}, B. 1988, \aap, 194, 335

\bibitem[{{Pinhas} {et~al.}(2019){Pinhas}, {Madhusudhan}, {Gandhi}, \&
  {MacDonald}}]{pinhasmadhusudhan2019}
{Pinhas}, A., {Madhusudhan}, N., {Gandhi}, S., \& {MacDonald}, R. 2019, \mnras,
  482, 1485

\bibitem[{{Pinhas} {et~al.}(2018){Pinhas}, {Rackham}, {Madhusudhan}, \&
  {Apai}}]{pinhasrackham2018}
{Pinhas}, A., {Rackham}, B.~V., {Madhusudhan}, N., \& {Apai}, D. 2018, \mnras,
  480, 5314

\bibitem[{{Piskunov} {et~al.}(1995){Piskunov}, {Kupka}, {Ryabchikova}, {Weiss},
  \& {Jeffery}}]{piskunov1995}
{Piskunov}, N.~E., {Kupka}, F., {Ryabchikova}, T.~A., {Weiss}, W.~W., \&
  {Jeffery}, C.~S. 1995, \aaps, 112, 525

\bibitem[{{Rackham} {et~al.}(2018){Rackham}, {Apai}, \&
  {Giampapa}}]{rackhamapai2018}
{Rackham}, B.~V., {Apai}, D., \& {Giampapa}, M.~S. 2018, \apj, 853, 122

\bibitem[{{Richard} {et~al.}(2012){Richard}, {Gordon}, {Rothman}, {Abel},
  {Frommhold}, {Gustafsson}, {Hartmann}, {Hermans}, {Lafferty}, {Orton},
  {Smith}, \& {Tran}}]{richardgordon2012}
{Richard}, C., {Gordon}, I.~E., {Rothman}, L.~S., {et~al.} 2012, \jqsrt, 113,
  1276

\bibitem[{{Robinson} {et~al.}(2017){Robinson}, {Fortney}, \&
  {Hubbard}}]{robinsonfortney2017}
{Robinson}, T.~D., {Fortney}, J.~J., \& {Hubbard}, W.~B. 2017, \apj, 850, 128

\bibitem[{{Rocchetto} {et~al.}(2016){Rocchetto}, {Waldmann}, {Venot}, {Lagage},
  \& {Tinetti}}]{rocchettowaldmann2016}
{Rocchetto}, M., {Waldmann}, I.~P., {Venot}, O., {Lagage}, P.-O., \& {Tinetti},
  G. 2016, \apj, 833, 120

\bibitem[{{Rothman} {et~al.}(2013){Rothman}, {Gordon}, {Babikov}, {Barbe},
  {Chris Benner}, {Bernath}, {Birk}, {Bizzocchi}, {Boudon}, {Brown},
  {Campargue}, {Chance}, {Cohen}, {Coudert}, {Devi}, {Drouin}, {Fayt}, {Flaud},
  {Gamache}, {Harrison}, {Hartmann}, {Hill}, {Hodges}, {Jacquemart}, {Jolly},
  {Lamouroux}, {Le Roy}, {Li}, {Long}, {Lyulin}, {Mackie}, {Massie},
  {Mikhailenko}, {M{\"u}ller}, {Naumenko}, {Nikitin}, {Orphal}, {Perevalov},
  {Perrin}, {Polovtseva}, {Richard}, {Smith}, {Starikova}, {Sung}, {Tashkun},
  {Tennyson}, {Toon}, {Tyuterev}, \& {Wagner}}]{rothman2013}
{Rothman}, L.~S., {Gordon}, I.~E., {Babikov}, Y., {et~al.} 2013, \jqsrt, 130, 4

\bibitem[{{Rothman} {et~al.}(2010){Rothman}, {Gordon}, {Barber}, {Dothe},
  {Gamache}, {Goldman}, {Perevalov}, {Tashkun}, \& {Tennyson}}]{rothman2010}
{Rothman}, L.~S., {Gordon}, I.~E., {Barber}, R.~J., {et~al.} 2010, \jqsrt, 111,
  2139

\bibitem[{{Samland} {et~al.}(2017){Samland}, {Molli{\`e}re}, {Bonnefoy},
  {Maire}, {Cantalloube}, {Cheetham}, {Mesa}, {Gratton}, {Biller}, {Wahhaj},
  {Bouwman}, {Brandner}, {Melnick}, {Carson}, {Janson}, {Henning}, {Homeier},
  {Mordasini}, {Langlois}, {Quanz}, {van Boekel}, {Zurlo}, {Schlieder},
  {Avenhaus}, {Beuzit}, {Boccaletti}, {Bonavita}, {Chauvin}, {Claudi}, {Cudel},
  {Desidera}, {Feldt}, {Fusco}, {Galicher}, {Kopytova}, {Lagrange}, {Le
  Coroller}, {Martinez}, {Moeller-Nilsson}, {Mouillet}, {Mugnier}, {Perrot},
  {Sevin}, {Sissa}, {Vigan}, \& {Weber}}]{samlandmolliere2017}
{Samland}, M., {Molli{\`e}re}, P., {Bonnefoy}, M., {et~al.} 2017, \aap, 603,
  A57

\bibitem[{{Sauval} \& {Tatum}(1984)}]{sauval_tatum1984}
{Sauval}, A.~J. \& {Tatum}, J.~B. 1984, \apjs, 56, 193

\bibitem[{{Schwarz} {et~al.}(2016){Schwarz}, {Ginski}, {de Kok}, {Snellen},
  {Brogi}, \& {Birkby}}]{schwarzginki2016}
{Schwarz}, H., {Ginski}, C., {de Kok}, R.~J., {et~al.} 2016, \aap, 593, A74

\bibitem[{{Schweitzer} {et~al.}(1996){Schweitzer}, {Hauschildt}, {Allard}, \&
  {Basri}}]{schweitzer1996}
{Schweitzer}, A., {Hauschildt}, P.~H., {Allard}, F., \& {Basri}, G. 1996,
  \mnras, 283, 821

\bibitem[{{Scott} \& {Duley}(1996)}]{scottduley1996}
{Scott}, A. \& {Duley}, W.~W. 1996, \apjs, 105, 401

\bibitem[{{Servoin} \& {Piriou}(1973)}]{servoinpiriou1973}
{Servoin}, J.~L. \& {Piriou}, B. 1973, Physica Status Solidi B Basic Research,
  55, 677

\bibitem[{{Sharp} \& {Burrows}(2007)}]{sharpburrows2007}
{Sharp}, C.~M. \& {Burrows}, A. 2007, \apjs, 168, 140

\bibitem[{{Smith} {et~al.}(1994){Smith}, {Robinson}, {Hyland}, \&
  {Carpenter}}]{smithrobinson1994}
{Smith}, R.~G., {Robinson}, G., {Hyland}, A.~R., \& {Carpenter}, G.~L. 1994,
  \mnras, 271, 481

\bibitem[{{Sneep} \& {Ubachs}(2005)}]{sneepubachs2005}
{Sneep}, M. \& {Ubachs}, W. 2005, \jqsrt, 92, 293

\bibitem[{{Snellen} {et~al.}(2014){Snellen}, {Brandl}, {de Kok}, {Brogi},
  {Birkby}, \& {Schwarz}}]{snellenbrandl2014}
{Snellen}, I.~A.~G., {Brandl}, B.~R., {de Kok}, R.~J., {et~al.} 2014, \nat,
  509, 63

\bibitem[{{Snellen} {et~al.}(2010){Snellen}, {de Kok}, {de Mooij}, \&
  {Albrecht}}]{snellendekok2010}
{Snellen}, I.~A.~G., {de Kok}, R.~J., {de Mooij}, E.~J.~W., \& {Albrecht}, S.
  2010, \nat, 465, 1049

\bibitem[{Sousa-Silva {et~al.}(2015)Sousa-Silva, Al-Refaie, Tennyson, \&
  Yurchenko}]{sousa-silvaal-refaie14}
Sousa-Silva, C., Al-Refaie, A.~F., Tennyson, J., \& Yurchenko, S.~N. 2015,
  Monthly Notices of the Royal Astronomical Society, 446, 2337

\bibitem[{{Sousa-Silva} {et~al.}(2014){Sousa-Silva}, {Hesketh}, {Yurchenko},
  {Hill}, \& {Tennyson}}]{sousa-silvahseketh14}
{Sousa-Silva}, C., {Hesketh}, N., {Yurchenko}, S.~N., {Hill}, C., \&
  {Tennyson}, J. 2014, \jqsrt, 142, 66

\bibitem[{{Sozzetti} {et~al.}(2015){Sozzetti}, {Bonomo}, {Biazzo}, {Mancini},
  {Damasso}, {Desidera}, {Gratton}, {Lanza}, {Poretti}, {Rainer}, {Malavolta},
  {Affer}, {Barbieri}, {Bedin}, {Boccato}, {Bonavita}, {Borsa}, {Ciceri},
  {Claudi}, {Gandolfi}, {Giacobbe}, {Henning}, {Knapic}, {Latham}, {Lodato},
  {Maggio}, {Maldonado}, {Marzari}, {Martinez Fiorenzano}, {Micela},
  {Molinari}, {Mordasini}, {Nascimbeni}, {Pagano}, {Pedani}, {Pepe}, {Piotto},
  {Santos}, {Scandariato}, {Shkolnik}, \& {Southworth}}]{sozzettibonomo2015}
{Sozzetti}, A., {Bonomo}, A.~S., {Biazzo}, K., {et~al.} 2015, \aap, 575, L15

\bibitem[{{Thalman} {et~al.}(2014){Thalman}, {Zarzana}, {Tolbert}, \&
  {Volkamer}}]{thalmanzarzana14}
{Thalman}, R., {Zarzana}, K.~J., {Tolbert}, M.~A., \& {Volkamer}, R. 2014,
  \jqsrt, 147, 171

\bibitem[{{Thalman} {et~al.}(2017){Thalman}, {Zarzana}, {Tolbert}, \&
  {Volkamer}}]{thalmanzarzana17}
{Thalman}, R., {Zarzana}, K.~J., {Tolbert}, M.~A., \& {Volkamer}, R. 2017,
  \jqsrt, 189, 281

\bibitem[{{Toon} \& {Ackerman}(1981)}]{toonackerman1981}
{Toon}, O.~B. \& {Ackerman}, T.~P. 1981, \ao, 20, 3657

\bibitem[{{Tremblin} {et~al.}(2015){Tremblin}, {Amundsen}, {Mourier},
  {Baraffe}, {Chabrier}, {Drummond}, {Homeier}, \&
  {Venot}}]{tremblinamundsen2015}
{Tremblin}, P., {Amundsen}, D.~S., {Mourier}, P., {et~al.} 2015, \apjl, 804,
  L17

\bibitem[{{van Boekel} {et~al.}(2012){van Boekel}, {Benneke}, {Heng}, {Hu},
  {Madhusudhan}, {Quanz}, {B{\'e}tr{\'e}mieux}, {Bouwman}, {Chen}, {Decin}, {de
  Kok}, {Glauser}, {G{\"u}del}, {Hauschildt}, {Henning}, {Jeffers}, {Jin},
  {Kaltenegger}, {Kerschbaum}, {Krause}, {Lammer}, {Luntzer}, {Meyer},
  {Miguel}, {Mordasini}, {Ottensamer}, {Rank-Lueftinger}, {Reiners},
  {Reinhold}, {Schmid}, {Snellen}, {Stam}, {Sun}, \&
  {Vandenbussche}}]{vanboekel2012}
{van Boekel}, R., {Benneke}, B., {Heng}, K., {et~al.} 2012, in Society of
  Photo-Optical Instrumentation Engineers (SPIE) Conference Series, Vol. 8442,
  Society of Photo-Optical Instrumentation Engineers (SPIE) Conference Series,
  1

\bibitem[{{Wakeford} \& {Sing}(2015)}]{wakefordsing2015}
{Wakeford}, H.~R. \& {Sing}, D.~K. 2015, \aap, 573, A122

\bibitem[{{Waldmann} {et~al.}(2015{\natexlab{a}}){Waldmann}, {Rocchetto},
  {Tinetti}, {Barton}, {Yurchenko}, \& {Tennyson}}]{waldmannrocchetto2015}
{Waldmann}, I.~P., {Rocchetto}, M., {Tinetti}, G., {et~al.} 2015{\natexlab{a}},
  \apj, 813, 13

\bibitem[{{Waldmann} {et~al.}(2015{\natexlab{b}}){Waldmann}, {Tinetti},
  {Rocchetto}, {Barton}, {Yurchenko}, \& {Tennyson}}]{waldmanntinetti2015}
{Waldmann}, I.~P., {Tinetti}, G., {Rocchetto}, M., {et~al.} 2015{\natexlab{b}},
  \apj, 802, 107

\bibitem[{{Wende} {et~al.}(2010){Wende}, {Reiners}, {Seifahrt}, \&
  {Bernath}}]{wendereiners2010}
{Wende}, S., {Reiners}, A., {Seifahrt}, A., \& {Bernath}, P.~F. 2010, \aap,
  523, A58

\bibitem[{{Woitke} {et~al.}(2016){Woitke}, {Min}, {Pinte}, {Thi}, {Kamp},
  {Rab}, {Anthonioz}, {Antonellini}, {Baldovin-Saavedra}, {Carmona}, {Dominik},
  {Dionatos}, {Greaves}, {G{\"u}del}, {Ilee}, {Liebhart}, {M{\'e}nard},
  {Rigon}, {Waters}, {Aresu}, {Meijerink}, \& {Spaans}}]{woitkemin2016}
{Woitke}, P., {Min}, M., {Pinte}, C., {et~al.} 2016, \aap, 586, A103

\bibitem[{{Yurchenko} {et~al.}(2018){Yurchenko}, {Al-Refaie}, \&
  {Tennyson}}]{yurchenkoalrefaie2018}
{Yurchenko}, S.~N., {Al-Refaie}, A.~F., \& {Tennyson}, J. 2018, \aap, 614, A131

\bibitem[{Yurchenko {et~al.}(2011)Yurchenko, Barber, \&
  Tennyson}]{yurchenkobarber2011}
Yurchenko, S.~N., Barber, R.~J., \& Tennyson, J. 2011, Monthly Notices of the
  Royal Astronomical Society, 413, 1828

\bibitem[{{Yurchenko} \& {Tennyson}(2014)}]{yurchenko2014}
{Yurchenko}, S.~N. \& {Tennyson}, J. 2014, \mnras, 440, 1649

\bibitem[{{Zhang} {et~al.}(2019){Zhang}, {Chachan}, {Kempton}, \&
  {Knutson}}]{zhangchachan2019}
{Zhang}, M., {Chachan}, Y., {Kempton}, E.~M.-R., \& {Knutson}, H.~A. 2019,
  \pasp, 131, 034501

\end{thebibliography}

\appendix

\section{\ptrad \ verification: very cloudy spectra}
\label{sect:app_very_cloudy}
\begin{figure*}[t!]
\centering
\includegraphics[width=1.\textwidth]{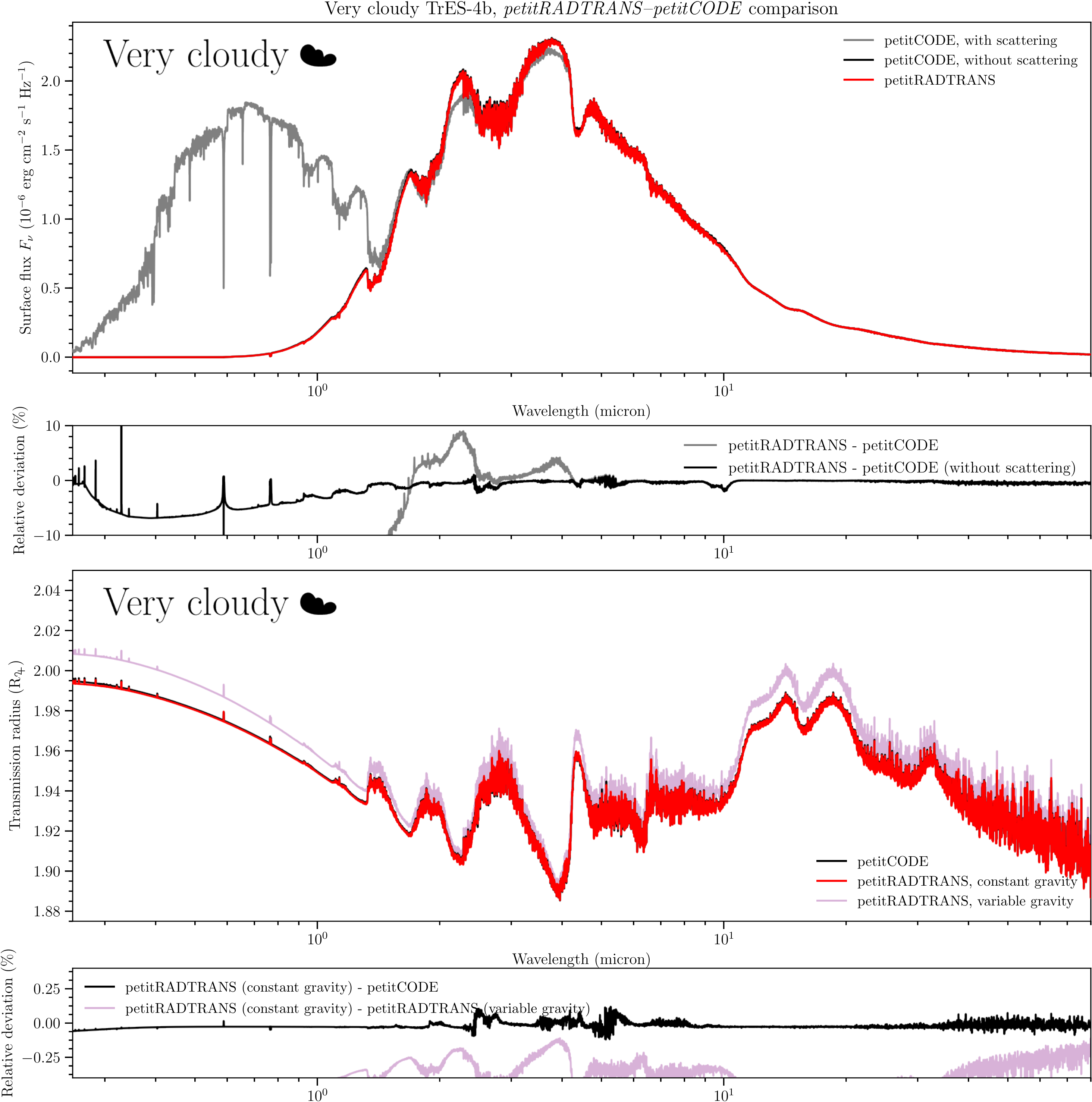}
\caption{Same as Figure \ref{fig:pt_rad_ver_clear}, but for the very cloudy case discussed in Section \ref{sect:cloudy_atmos}.}
\label{fig:pt_rad_ver_very_cloudy}
\end{figure*}

Analogous to Figure \ref{fig:pt_rad_ver_cloudy}, Figure \ref{fig:pt_rad_ver_very_cloudy} shows the comparison between \petit and \ptrad, but for an even cloudier hot Jupiter atmosphere, where the mass fraction of the clouds are only capped for values larger than $10^{-2}\cdot Z_{\rm P}$. This case is also discussed in Section \ref{sect:cloudy_atmos}.

\end{document}